\documentclass{emulateapj}
\usepackage[colorlinks,urlcolor=blue,citecolor=blue,linkcolor=blue]{hyperref} 
\usepackage{graphicx,natbib}
\usepackage[space]{grffile}
\usepackage{latexsym}
\usepackage{amsfonts,amsmath,amssymb}
\usepackage{url}
\usepackage[utf8]{inputenc}
\usepackage{fancyref}
\usepackage{hyperref}
\usepackage{multirow}
\hypersetup{colorlinks=false,pdfborder={0 0 0},}

\DeclareMathOperator{\erfinv}{erfinv}

\begin{document}

\title{A Millisecond Interferometric Search for Fast Radio Bursts \\ with the Very Large Array}
\shorttitle{Millisecond Interferometric Search for FRBs}
\shortauthors{Law et al.}

\author{Casey J. Law\altaffilmark{1}}
\author{Geoffrey C. Bower\altaffilmark{2}}
\author{Sarah Burke-Spolaor\altaffilmark{3,4}}
\author{Bryan Butler\altaffilmark{4}}
\author{Earl Lawrence\altaffilmark{5}}
\author{T. Joseph W. Lazio\altaffilmark{3}}
\author{Chris A. Mattmann\altaffilmark{3}}
\author{Michael Rupen\altaffilmark{6}}
\author{Andrew Siemion\altaffilmark{1}}
\author{Scott VanderWiel\altaffilmark{5}}
\altaffiltext{1}{Dept of Astronomy and Radio Astronomy Lab, Univ. of California, Berkeley, CA}
\altaffiltext{2}{Academia Sinica Institute of Astronomy and Astrophysics, 645 N. A'ohoku Place,Hilo, HI 96720, USA; \href{gbower@asiaa.sinica.edu.tw}}
\altaffiltext{3}{Jet Propulsion Laboratory, California Institute of Technology, Pasadena, CA}
\altaffiltext{4}{National Radio Astronomy Observatory, Socorro, NM}
\altaffiltext{5}{Los Alamos National Laboratory, Los Alamos, NM}
\altaffiltext{6}{NRC Herzberg, Penticton, BC, Canada}

\begin{abstract}
We report on the first millisecond timescale radio interferometric search for the new class of transient known as fast radio bursts (FRBs). We used the Very Large Array (VLA) for a 166-hour, millisecond imaging campaign to detect and precisely localize an FRB. We observed at 1.4 GHz and produced visibilities with 5 ms time resolution over 256 MHz of bandwidth. Dedispersed images were searched for transients with dispersion measures from 0 to 3000 pc cm$^{-3}$. No transients were detected in observations of high Galactic latitude fields taken from September 2013 though October 2014. Observations of a known pulsar show that images typically had a thermal-noise limited sensitivity of 120 mJy beam$^{-1}$ ($8\sigma$; Stokes I) in 5 ms and could detect and localize transients over a wide field of view. Our nondetection limits the FRB rate to less than $7\times10^4$ sky$^{-1}$ day$^{-1}$ (95\% confidence) above a fluence limit of 1.2 Jy-ms. Assuming a Euclidean flux distribution, the VLA rate limit is inconsistent with the published rate of Thornton et al. We recalculate previously published rates with a homogeneous consideration of the effects of primary beam attenuation, dispersion, pulse width, and sky brightness. This revises the FRB rate downward and shows that the VLA observations had a roughly 60\% chance of detecting a typical FRB and that a 95\% confidence constraint would require roughly 500 hours of similar VLA observing. Our survey also limits the repetition rate of an FRB to 2 times less than any known repeating millisecond radio transient.

\end{abstract}

\bibliographystyle{apj}

\section{Introduction}
Large radio pulsar surveys with single-dish telescopes have revealed a new class of transient known as a ``fast radio burst'' \citep[FRB;][]{2013Sci...341...53T}.
Several FRBs have now been detected in multi-beam pulsar surveys at the Parkes and Arecibo obseratories \citep{2007Sci...318..777L, 2011MNRAS.415.3065K, 2014ApJ...790..101S, 2014arXiv1412.1599R, 2014arXiv1412.0342P}. The key feature of the FRB is its large dispersion measure (DM), which ranges from 300 to 1100 pc cm$^{-3}$ and is typically an order of magnitude larger than expected from the Galaxy. One explanation for this large DM is that the bursts are dispersed by low density ionized plasma in the intergalactic medium (IGM), implying that they originate at distances up to and beyond redshifts of 1.

Models for the origin of FRBs must account for radio luminosities higher than $10^{12}$ Jy-kpc$^2$, far beyond that of Galactic neutron star transients \citep{2003ApJ...596..982M}. Despite their unusual luminosity, their occurrence rate is $10^4$ sky$^{-1}$ day$^{-1}$ for brightnesses of $\sim3$\ Jy-ms \citep{2013Sci...341...53T}, which is about as frequent as core-collapse supernovae within roughly 1 Gpc. Several kinds of cataclysmic events have been proposed to produce FRBs, such as the births of degenerate objects, the evaporation of primordial black holes \citep{1977Natur.266..333R}, and mergers of binary degenerate objects \citep{2014A&A...562A.137F,2013ApJ...776L..39K}. Despite dedicated searches, no FRB has been associated with a GRB \citep{2012ApJ...757...38B,2014ApJ...790...63P}.

FRBs have huge potential in understanding the intergalactic medium and measuring cosmological parameters. The dispersion of an extragalactic transient measures the electron column density, a good proxy for baryonic mass. Detailed models of the Galaxy's electron density have been built by associating dispersion with sources of known distance \citep{2002astro.ph..7156C} and FRBs could allow us to do the same for the IGM \citep{2013ApJ...776..125M}. Even in the local universe, the dispersion measure of any pulses from outside our Galaxy would measure the baryon content in the diffuse halo to potentially solve the ``missing baryon problem'' \citep{2007ARA&A..45..221B,2013ApJ...762...20F}. At cosmological distances, FRBs could test models for dark energy in a new way \citep{2014ApJ...783L..35D}.

The story of the FRB is complicated by the simultaneous discovery of a new class of terrestrial interference known as perytons \citep{2011ApJ...727...18B}. Perytons are impulsive radio transients with a width of tens of ms and an apparent DM of a few hundred, partially overlapping with characteristics expected of extragalactic radio transients. Perytons tend to be detected in the month of July and have arrival times that seem to correlate with the integer second, suggesting that they are man made \citep{2012MNRAS.420..271K}. However, they are distinguished from FRBs by an apparent degree-scale extent on the sky (seen in all beams of multi-beam receivers), non-uniform spectral structure, tens of milliseconds pulse widths, and small range of apparent DMs \citep{2012MNRAS.425.2501B}. To complicate the story further, the first FRB \citep{2007Sci...318..777L} appears to share the properties of both perytons (high brightness, similar DM) and FRBs (small spatial extent and pulse width). The recent detection of FRBs by the Arecibo Observatory shows that FRBs are not unique to the environment of discovery telescope \citep{2014ApJ...790..101S}. Perytons have been detected at observatories in Switzerland \citep{2014ApJ...795...19S} and excluded at the Allen Telescope Array in northern California \citep{2012ApJ...744..109S}, which shows that they have a wide ranging, but distinctly terrestrial distribution. \citet{2014ApJ...797...70K} attempt to unify both classes of radio transient as a terrestrial phenomenon seen in the near and far-field optics regimes of the radio telescopes. Those authors note that the VLA can test this model, since its far-field limit is near the moon. An interferometric VLA search would detect FRBs, but would not be sensitive to perytons.

An interferometer like the VLA is needed not only to test whether FRBs are astrophysical, but to fulfill their scientific potential. Single-dish radio telescopes localize transients with a precision of order 10\arcmin. This is too coarse to uniquely associate an FRB with a multiwavelength counterpart \citep[unless it is also variable at other wavelengths;][]{1997Natur.387..878M,2014arXiv1412.0342P}. Furthermore, since single-dish telescopes have a location-dependent sensitivity and spectral response, not knowing the location of the transient within that region makes it difficult to measure its luminosity or spectrum. By contrast, the VLA can find a transient over wide field of view, localize it to arcsecond precision, and unambiguously measure its properties \citep{2012ApJ...760..124L}.

This paper describes the implementation of an imaging search at the Karl G.~Jansky Very Large Array (VLA), capable of achieving arcsecond localization of FRBs, and the initial results from a 166-hour survey. This effort makes use of the new high data rate capabilities of the VLA to produce 5-millisecond visibilities, which we dedisperse and image to search for transients with a parallelized pipeline. In Section \ref{daq}, we describe the data acquisition and survey parameters. Section \ref{proc} describes the data management, parallelized transient search software, and computing systems. Section \ref{cand} presents our analysis of the most significant events, all of which are attributed to thermal noise or interference. The data quality and survey sensitivity are discussed in Section \ref{qual}. We present our estimate of the upper limit on the rate of FRBs in Section \ref{rate}, including a new homogeneous definition of flux limit of previous surveys. 

\section{Data Acquisition and Survey Specification}
\label{daq}
Throughout 2012, our team commissioned the VLA to collect millisecond-scale correlated data products to be imaged and searched for transients \citep{2012ApJ...760..124L}. The spatial information measured by interferometers comes at the cost of higher data rates and computational burden. Ideally, we would search with maximal bandwidth and a time resolution faster than 1 ms, the temporal width of the narrowest FRBs. The highest data rate throughput for the VLA correlator is roughly 285 MB s$^{-1}$ or 1 TB hour$^{-1}$. That allowed us to use an integration time of 5 ms (modestly underresolving the pulse width) with a bandwidth of 256 MHz and 256 spectral channels. This time and spectral resolution are well matched when searching for highly-dispersed transients.

Observations used the full array, of which between 0 and 3 antennas were removed for substandard performance. Data were acquired in two orthogonal circular polarizations, each with two spectral windows covering frequencies from 1268 to 1524 MHz, similar to the discovery observations made with Parkes \citep[e.g.,][]{2007Sci...318..777L}. We used a band centered at 1396 MHz with a bandwidth of 232 MHz to avoid band edge effects.

The selection of pointing locations was guided by strategic issues. The first criterion was to observe at high Galactic latitudes with low brightness temperature to avoid the influence of Galactic dispersion and scattering. Second, we preferred fields at low elevations in order to reduce the number of independent pixels in our images and corresponding computing burden (see \S \ref{proc}). Third, we avoided declinations near --5\arcdeg, which are heavily affected by radio frequency interference (RFI). For scheduling flexibility, we defined several locations over a range of sidereal times with relatively low demand (see Table \ref{fields}). One field is centered on the position of FRB120127 \citep{2013Sci...341...53T}, two are in deep, multi-wavelength survey fields (COSMOS and Chandra Deep Field South), two are pointed at faint pulsars (as a secondary check of detection reliability), and the other three are otherwise unconstrained.

Observations took place in two broad campaigns from September 2013 through January 2014 and from June 2014 through October 2014 (see Tables \ref{fields} and \ref{processing}). The first campaign observed for 76 hours (total time on sky) in CnB and B configurations. The second campaign observed for 124 hours in A, D, DnC, and C configurations. Observations were also made during the antenna reconfiguration periods, so antenna availability and baseline lengths can vary from day to day. The total scheduled time was 201 hours with an on-target (ignoring time on calibrators and moving telescope) observing efficiency of 82\%. This gave us 166 hours of time on target fields that was searched for FRBs.

Observing sessions lasted from 45 minutes to 6 hours. A typical observation lasted 2 hours \footnote{Two hours of observing in this mode produces 2 TB of data, a substantial fraction of the amount of data that can be transferred off site in a day. As a result, observations were carefully scheduled to avoid affecting normal VLA operations.} and was composed of 50, 2-minute scans at the target field with gain calibration scans interspersed every 20-30 minutes. Roughly once per week of observing, we also observed a flux calibrator (either 3C48 and 3C147) and the pulsar B0355+54 as an end-to-end test of our transient detection system.

\begin{table}
\caption{Survey Fields}
\footnotesize
\centering
\begin{tabular}{l|cc|cc|c}
\hline
Field       & RA          & Dec   & Lon. & Lat.     & Time \\
            & \multicolumn{2}{|c|}{(J2000)}  & \multicolumn{2}{|c|}{(Galactic; deg)} & (hrs) \\ \hline
RA02        & 2:27:53  &  +9:13:24 & 159.0 & --46.8    & 26.25 \\
CDF-South   & 3:32:28  & --27:48:30  & 223.6 & --54.4  & 4  \\
RA05        & 5:04:37  & --30:50:00  & 233.0 & --35.3   & 16 \\
COSMOS      & 10:00:29 &  +2:12:21    & 236.8 & 42.1   & 24 \\
RA12          & 12:00:7  &  +5:53:10  & 270.7 &   65.5 & 4.75 \\
FRB120127     & 23:15:00   & --18:25:00  & 49.3 & --66.2   & 40 \\
PSR J2013-0649 & 20:13:18 & --6:49:5 & 36.2 & --21.3 & 79.5 \\
PSR J2248-0101 & 22:48:27 & --1:1:48 & 69.3 & --50.6 & 6.5 \\ \hline
\end{tabular}
\label{fields}
\end{table}

\section{Transient Search Processing}
\label{proc}
We developed a parallelized software system to search visibility data for transients\footnote{Code is available at \url{http://github.com/caseyjlaw/tpipe}.}. Our software extends single-dish time-domain techniques (e.g., dedispersion) to visibility-type data and integrates it into a pipeline that performs parallelized interferometric imaging. We wrote this system in Python to take advantage of the the substantial code base for interferometric data management that exists within the NRAO CASA software package \citep{2007ASPC..376..127M}. Key custom functions, such as dedispersion and imaging, are accelerated with Cython, a compiled form of Python\footnote{See \url{http://cython.org}.}.

We applied calibration products generated by either CASA or the on-line TelCal system. The TelCal system ran phase calibration for all gain calibrators observed by the VLA and produces a solution for each scan, antenna, polarization, and spectral window. Offline, we also used CASA to flag data and create bandpass-corrected gain solutions for each observation. Roughly once per week, we observed a flux calibrator and calculated flux calibrated gain corrections. We ran a script to compare the image quality of the CASA and TelCal solutions and applied the calibration products that produced the highest S/N detections of the calibrators. For all but 9 observations, the CASA solutions were judged best; when TelCal solutions were used, the image flux scale had arbitrary units, but the analysis was otherwise identical.

The transient detection pipeline uses a hybrid parallelization model. Each node of a cluster runs an independent search on a single, 2-minute scan of data. Since each node is entirely independent, this level of parallelization is 100\% efficient. On each node, we use the Python \emph{multiprocessing} library to create two threads for data preparation and processing. The processing thread launches many more threads (fully utilizing available cores) to parallelize the dedispersion and imaging steps. The efficiency of this parallelism is sensitive to the size of the data being read, the image size, and other factors, as described below.

\begin{figure}[htb]
\begin{center}
\includegraphics[width=0.9\columnwidth]{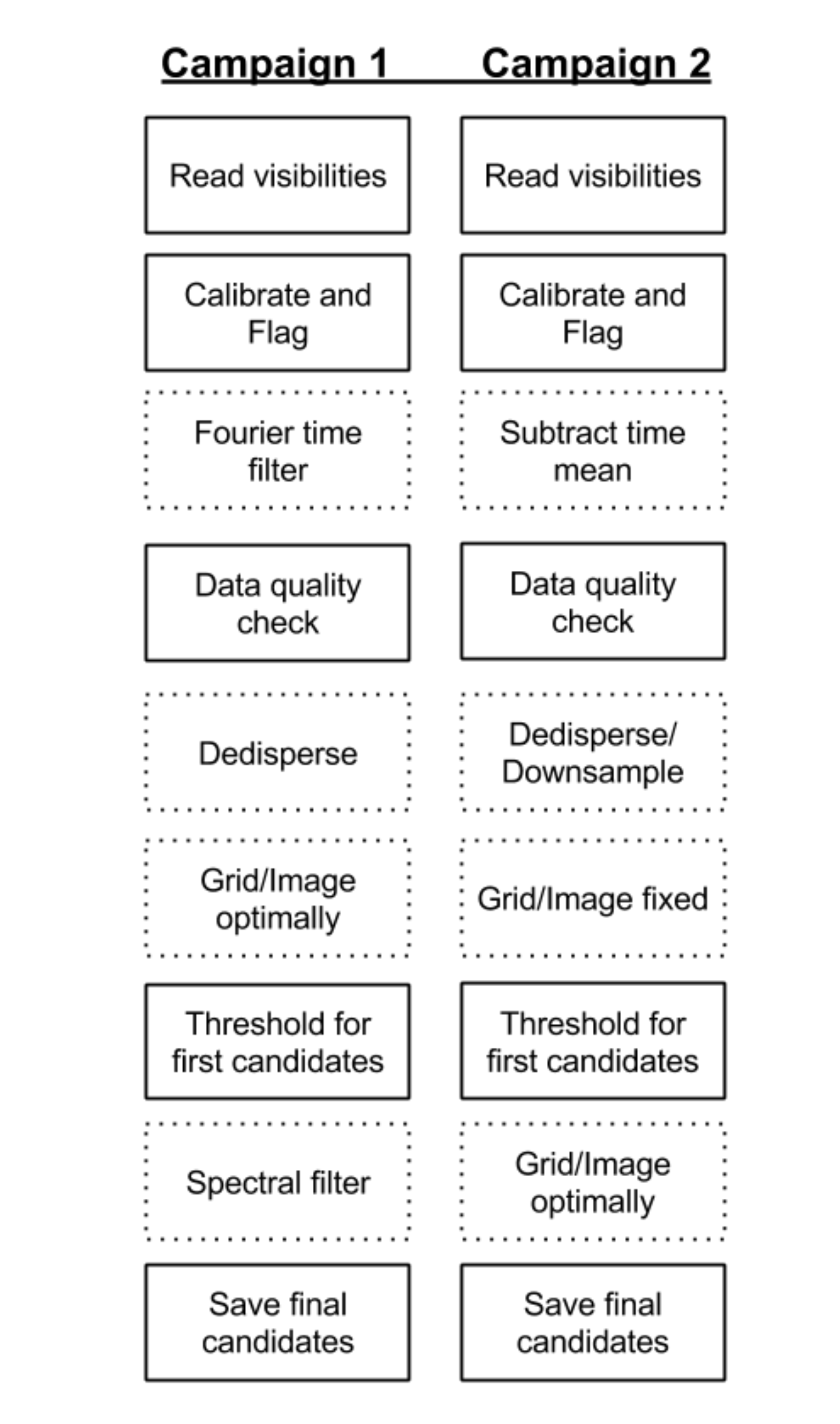}
\caption{Schematic view showing the progression of operations in the transient search pipeline, ordered from top to bottom. Dotted boxes show aspects of the pipeline that changed between Campaigns 1 and 2.
\label{procschem}}
\end{center}
\end{figure}

The transient search pipeline is designed to find transient point-like sources in images made from dedispersed data. While algorithms involving interferometric closure quantities are known to be faster \citep{2012ApJ...749..143L}, we required maximal sensitivity to have a reasonable chance at detecting an FRB. Figure \ref{procschem} shows the pipeline schematically and how it changed between the first and second observing campaigns. In all cases, the pipeline prepares data by dynamically applying calibration, flagging bad data, and removing non-variable sources by subtracting the mean visibility in time. The visibilities are then dedispersed, imaged, and images are searched for bright point sources throughout a 1\arcdeg\ field. That field covers twice the full-width half-max (FWHM) of the primary beam at 1.4 GHz. In detail, the pipeline stages are:
\begin{enumerate}
 \item {\bf Read data.} Due to memory limitations, we typically read data in segments shorter than 200 integrations, equivalent to 1 second. Data are read into shared memory to allow communication between reading and processing threads.
 \item {\bf Apply calibration.} A custom function parses either TelCal or CASA calibration products and applies the solution that is nearest in time to the segment. Antennas/polarizations with very low gain amplitudes are flagged at this stage.
 \item {\bf Flag bad data.} We use four criteria to flag bad data. First, we flag baselines with strong spectral periodicity, which is indicative of Gibbs ringing from strong RFI. Next, we flag individual channels and integrations with very large amplitude deviations from the median value. Third, we flag antennas and polarizations with very large amplitudes. Finally, we apply a uniquely interferometric flagging criterion, which is based on the standard deviation of the complex visibility over all baselines for a given channel, integration, and polarization. Many kinds of RFI affect a subset of the array, so this quantity identifies it, even if it only affects a single channel and integration. The flagging fraction ranged from 7\% to 40\% with a median value of about 15\%.
 \item {\bf Subtract mean visibility in time.} To remove constant sources and assure a zero-mean visibility distribution, we subtracted a local estimate of the mean visibility (on real-imaginary values) for each segment. Background subtraction is done by either a Fourier filter or a simple mean subtraction. For the first campaign, we subtracted the background by convolving visibilities with a time-domain convolutional kernel shaped as a zero-mean delta function. The function has a single integration of height 1 surrounded by a bin with a value of zero, and then $N$\ integrations with value $-1/N$. We use a background window of 5, so the total width of the kernel is 13 integrations or 65 milliseconds. The convolution is applied for each segment and the filter wraps at the time boundaries. For a segment length $\tau=1$\ seconds, the largest visibility phase change within the segment due to earth rotation is $\phi=2\pi*(b/D)*(\tau/24 \rm{hr}) \approx 5$\arcdeg. In the second campaign, we opted for a simpler approach --- subtraction of the mean visibility in time over the segment.
 \item {\bf Measure data quality.} Each prepared segment of data (mostly 200 integrations long) is summarized with data quality indicators. We track data quality by measuring the standard deviation of the visibilities, the fraction of data flagged, and the standard deviation of pixels in a single image for DM$=0$. The data are then handed off for processing and the reading loop starts preparing the next segment of data.
 \item {\bf Dedisperse and downsample visibilities.} A new thread is launched to dedisperse visibilities for each DM. We used 119 DM trials ranging from 0 to 3000 pc cm$^{-3}$. This DM grid was designed to allow at most 25\% sensitivity loss between DM trials, as is typical for such algorithms \citep{2010MNRAS.409..619K}. Since the dispersion delay across the band is large relative to the segment size, we accumulate dedispersed data over as many segments as are needed to cover the dispersive delay. In the first campaign, we downsampled (summed) in time for DM greater than 1850 pc cm$^{-3}$, where the dispersion shift was larger than 1 integration per channel. Since the Fourier time filter used in the first campaign has a width of 1 integration, sensitivity to high-DM signals is suppressed by roughly 25\% relative to the theoretical maximum. For the second campaign, we downsampled by summing adjacent integrations for all DM values; this search covered timescales of 1, 2, 4, and 8 integrations (5, 10, 20, and 80 ms). This function is accelerated in Cython.
 \item {\bf Grid visibilities.} The same thread then places the dedispersed visibilities on a regular \emph{uv} grid. Visibilities for both polarizations are gridded with a tophat convolutional kernel with ``natural'' weighting (maximally sensitive) to produce Stokes I images. A single \emph{uv} grid definition is defined for the middle integration of the segment. For a 200-integration-long segment, the fixed grid introduces errors due to the Earth's rotation on the scale of at most 1 second, equal to $4\lambda$ and always much smaller than the \emph{uv} grid cell. The function is written in Cython, which allows us to vectorize gridding in time and polarization. The transient search used a grid cell size of 58 lambda, which introduces a grid beam (grid-based decorrelation) $1/\Delta u\approx1$\arcdeg\ (twice the primary beam FWHM at 1.4 GHz). For the first campaign, the number of pixels in the grid is defined to include all visibilities (rounded up to the nearest multiple of 2 or 3 for efficiency of the Fast Fourier Transform; FFT), which produces two image pixels per synthesized beam size. In the second campaign, we used two-stage imaging, where an initial search was made with 512 pixels, which ignores data on the longest baselines and reduces computational demand in A configuration. All candidates from the first stage were imaged a second time with all data. While antennas were being moved from A to D configuration, the fixed \emph{uv} grid included anywhere from 30\% to 70\% of the data (details in \S \ref{cand}).
 \item {\bf Image.} We then perform a 2d FFT to form an image of each integration. No primary beam correction is applied, so the image noise distribution is uniform, but sensitivity is not. For performance reasons, images are not deconvolved to remove the effect of the point spread function; this biases the apparent S/N, but only for S/N values higher than our threshold. The peak and standard deviation of the pixel values in each image is calculated. If the S/N of the peak is larger than a predefined threshold ($6.5\sigma$\ in the first campaign and $6.0\sigma$\ in the second campaign), then information about the candidate is returned. The threshold is chosen to catch the tail of the thermal noise distribution of candidates. The final image pixel scale is given in Table \ref{processing}.
 \item {\bf Filter candidates.} In the first campaign, we phase the visibilities to the location of the each candidate and measure its spectrum. The spectral modulation \citep[normalized standard deviation over channels;][]{2012ApJ...748...73S} is used to reject candidates with most of their power in a narrow range of frequencies (i.e., RFI). This filter has a tunable parameter that we set to reject obvious RFI, but include all known pulses from a test scan of a pulsar. In some cases, this filter reduced the number of RFI-generated candidates by an order of magnitude, which greatly simplified candidate inspection. In the second campaign, we improved real-time RFI flagging, so the spectral modulation filter was removed. The two-stage imaging meant that any candidate needed to have a significance greater than $6.0\sigma$ in images made with a subset and with all data.
 \item {\bf Save candidate info.} If a candidate passes these tests, then a host of information about the candidate is saved to disk. This information is later used to visualize the candidate during manual inspection.
\end{enumerate}

\begin{table*}
\caption{Processing version and imaging parameters for each field and configuration}
\centering
\begin{tabular}{l|ccccc}
Campaign               & Field     & Configuration & Time & Image size & Pixel size \\
                       &           &               & (hours) & (pixels) & (arcseconds) \\ \hline
\multirow{7}{*}{First} & RA02      & B & 2 & (1458, 1458) & (2.4, 2.4) \\
                       & CDF-South & B & 4 & (1728, 864) & (2.1, 4.1) \\
                       & RA05      & CnB & 10 & (576, 512)  & (6.2, 7.0) \\
                       & RA05      & B   & 6 & (1728, 768) & (2.1, 4.7) \\
                       & COSMOS    & B   & 10 & (1458, 1296) & (2.4, 2.7) \\
                       & RA12      & B   & 4  & (1752, 1361) & (2.1, 2.6) \\
                       & FRB120127 & B   & 40 & (1944, 1024) & (1.9, 3.5) \\ \hline  
\multirow{12}{*}{Second}& RA02  & A   & 4 & (5184, 5374)  & (0.7, 0.8) \\
                       & RA02  & D   & 10.5 & (192, 128)  & (19, 28) \\
                       & RA02  & DnC &  9.75 & (192, 324) & (19, 11) \\
                       & COSMOS & A  &  2    & (5184, 4374) & (0.7, 0.8) \\
                       & COSMOS & D  &  0.75 & (192, 144) & (19, 25) \\
                       & COSMOS & DnC & 10.5 & (192, 324) & (19, 11) \\
                       & COSMOS & C & 0.75   & (216, 384) & (17, 9.4) \\
                       & RA12   & DnC & 0.75 & (192, 324) & (19, 11) \\
                       & PSR J2013-0649 & A & 48    & (5184, 4374) & (0.7, 0.8) \\
                       & PSR J2013-0649 & D & 16.5  & (192, 128) & (19, 28) \\
                       & PSR J2013-0649 & DnC & 15  & (162, 288) & (22, 13) \\
                       & PSR J2248-0101 & D & 6.5   & (192, 128)  & (19, 28) \\ \hline
\end{tabular}
\tablecomments{The ``first'' and ``second'' campaigns refer to observations from September 2013 and January 2014 and from June 2014 and October 2014, respectively. The parameters of the transient search changed slightly between these campaigns, as described in \S \ref{proc}. Antenna configuration is identified by longest baseline in the array during the observation. \label{processing}}
\end{table*}

For contemporary multi-core CPU architectures (e.g., Intel Xeon, 16 cores, 64 GB memory), the processing time and memory footprint were dominated by the FFT stage of imaging in configurations larger than D. Observations were made in almost every antenna configuration, with maximum baselines from 1 to 31 km and \emph{uv} grids extending from 128 to 5184 pixels on a side. As noted earlier, the second campaign made the images in two stages with the first stage imaging using a fixed grid of 512 pixels. For these images, the processing pipeline running on 14 nodes at the AOC can search two hours of data in 8 hours, equivalent to 660 images per second per node or roughly $10^4$ images per second. A transient search of 2 hours of data produced $170\times10^6$ images and was completed within 1 day of the completion of observing.

The threshold to save candidate information was set low enough to trigger false positives from thermal noise, which allowed us to use the rate of false positives as a simple test of functionality and to measure the noise properties based on the S/N distribution. As our experience with the data grew, our RFI flagging algorithms became effective enough to eliminate nearly all RFI-generated false positives (as detailed below in \S \ref{cand}). Thus, we could predict the number of candidates for a given threshold by assuming that each pixel, integration, and DM trial is independent and equally likely to generate a thermal-noise candidate. We tested this idea with simulations of our transient detection pipeline with pure thermal-noise data and showed that there was a small amount of correlation between neighboring pixels (see \S \ref{cand}).

\begin{figure*}[h!]
\begin{center}
\includegraphics[width=2\columnwidth]{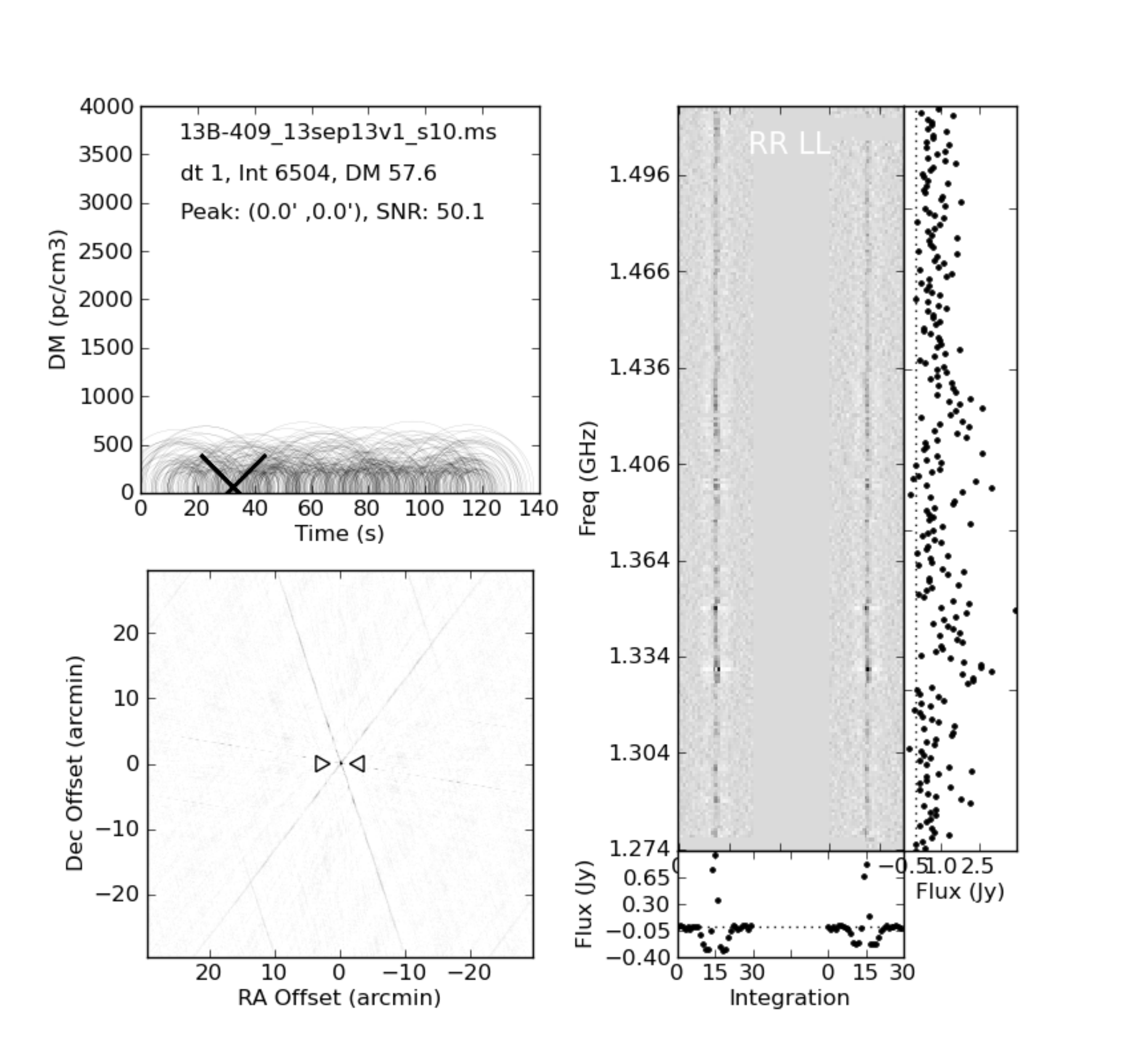}
\caption{Example candidate plot for a pulse detected in the first campaign from pulsar B0355+54. The text in the top left panel describes the candidate and the points show the DM-time distribution of all candidates cluster at the known pulsar DM of 57 pc cm$^{-3}$. The circle size represents the candidate S/N and a cross shows the candidate data shown in other panels of the plot. The background shows the DM-time distribution of all candidates in this 2-minute scan of data; the bold cross shows the selected candidates and circles show all others. The bottom left panel shows the image with the candidate location highlighted with triangles. Note that the point spread function is not deconvolved, so the image of this bright pulse shows the beam pattern also. The right panel shows the dedispersed pulse spectrogram (frequency vs time) for 15 integrations around the pulse. The two polarizations (RR and LL) are shown separately in the spectrogram, but are summed for the pulse spectrum from the central channel, which is shown furthest right. Strong scintillation enhances the pulsar brightness at some frequencies. At the bottom of plot, we take the mean over channels to show the dedispersed time series for the RR and LL polarizations. The negative flux at times adjacent to the pulse are produced by the Fourier-domain background filter (used in first campaign only; see \S \ref{proc}). \label{candplot}}
\end{center}
\end{figure*}

The principle product of the pipeline was a set of files containing information needed to reproduce the candidates, such as time, DM, and pulse width. The candidate files also included the S/N and location of the candidate within the image, which allowed us to generate summary plots that help identify interesting candidates or do further RFI flagging. Finally, the good candidates (defined below) were reproduced and summarized in a plot like that shown in Figure \ref{candplot}.

The bulk of the data processing was done using up to 15 nodes of the compute cluster at the NRAO Array Operations Center (AOC) in Socorro. Roughly one quarter of the data were processed at the ``darwin'' cluster at the Los Alamos National Lab. Finally, 76 TB of data were transferred to the National Energy Research Scientific Computing Center in Oakland, CA for archiving and in anticipation of further processing. The data are also publicly available on the NRAO archive under project codes 13B-409 and 14A-425.

\section{Candidate Analysis}
\label{cand}
The threshold for candidate analysis was set low enough to capture thermal-noise false positives. This allows us to use the candidate S/N distribution to measure the thermal-noise rate and identify high-S/N candidates that deserve manual inspection. To do this, we use a ``normal quantile plot'', where the observed S/N is compared to the expected S/N given the rate at which events of a given S/N are detected \citep{chambers1983graphical}. The normal quantile plot is useful because it allows us to easily look for deviations from thermal noise without being biased by the varying number of detections between observations due to changes in image size and duration. It is calculated by sorting the observed candidate S/Ns and calculating the quantile of each event in the sorted list:
\begin{equation}
q(i) = (n_{\rm{trials}} + 1/2 - i)/n_{\rm{trials}}
\end{equation}
\noindent where $i$ is the event location in the sorted list (1 is highest S/N candidate) and $n_{\rm{trials}}$ is the total number of independent samples. In this case the number of trials is the product of the number of pixels in each image, number of integrations, and the number of DM trials. The rate for a given S/N can be scaled to an expected S/N by assuming it is drawn from a normal distribution with a given number of trials:
\begin{equation}
Z(i) = \sqrt{2} \erfinv{(2*q(i) - 1)}
\end{equation}

\begin{figure}[htb]
\begin{center}
\includegraphics[width=\columnwidth]{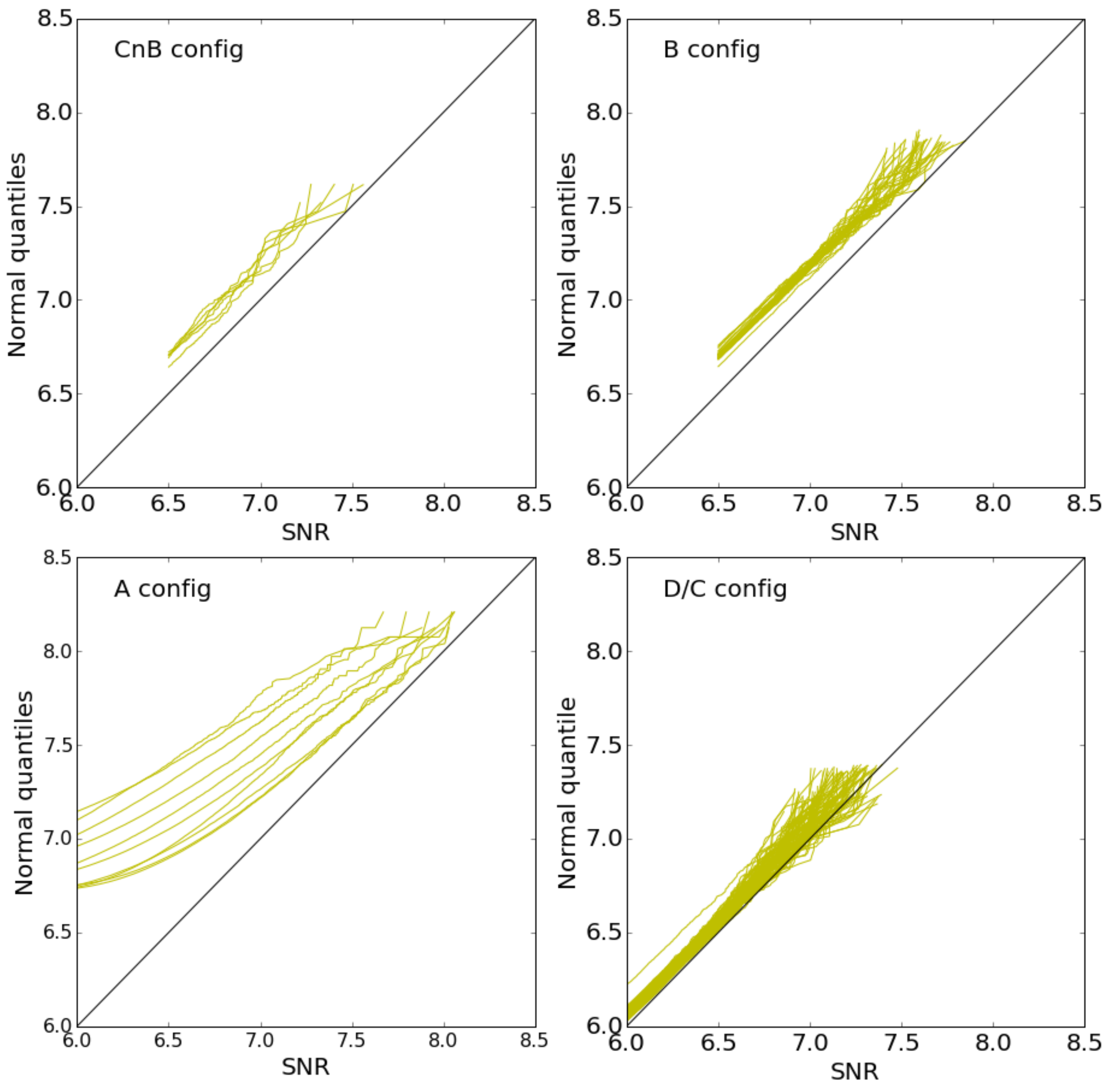}
\caption{The four panels show the cumulative event rate to estimate the expected S/N assuming a normal distribution (a.k.a., the standard normal quantile). Each panel shows the data from a different VLA antenna configuration, with each line showing the S/N distribution of a single observation. RFI and known pulsar pulses have been removed. The black line shows the expected trend for total independence of all trials/pixels/integrations and the most significant events are seen toward the right.
\label{normprob}}
\end{center}
\end{figure}

Figure \ref{normprob} shows quantiles of the upper tails of the observed S/N data plotted against the theoretical quantiles of a thermal noise distribution. Assuming independent Gaussian pixels and DM trials, the data should follow the black line. The panels show different configurations and each line is a different observation (typically 2 hours). The important feature of these plots is the consistency of the lines for each observation within each panel (with the exception of A configuration; see below). This suggests that the upper tail of the thermally-induced noise distribution is not varying from observation to observation (as a result, for example, of some variation in the array). This is good evidence that our choice of constant threshold trigger should be effective.

There are several reasons that the independent Gaussian pixel model does not hold. First, pixels are correlated within an image because the $uv$\ grid is only partially populated with visibilities. Next, each DM trial partially overlaps others in frequency-time space, so these trials are also correlated. Observations from D, DnC, and C configurations produced candidates directly from bright pixels in images with no further filtering. The 'D/C' panel shows that the independent pixel model is only slightly conservative; conservative means that the thermal S/N values are not as large as would be produced by independent pixels so the false detection rate above, e.g., $8\sigma$ is less than $6.2\times10^{-16}$ per pixel as determined from a Gaussian distribution.

Candidates in the CnB and B configurations were also required to pass the spectral modulation filter. This second stage culling produces fewer large S/N candidates, shifting the corresponding lines in Figure \ref{normprob} to the left and rendering an $8\sigma$ threshold even more conservative. The A configuration also used two stages of filtering but the effect of second stage imaging lessened over the course of the campaign as additional antennas were moved into their compact locations and added short baselines for the first stage of imaging.

\begin{figure}[htb]
\begin{center}
\includegraphics[width=1\columnwidth]{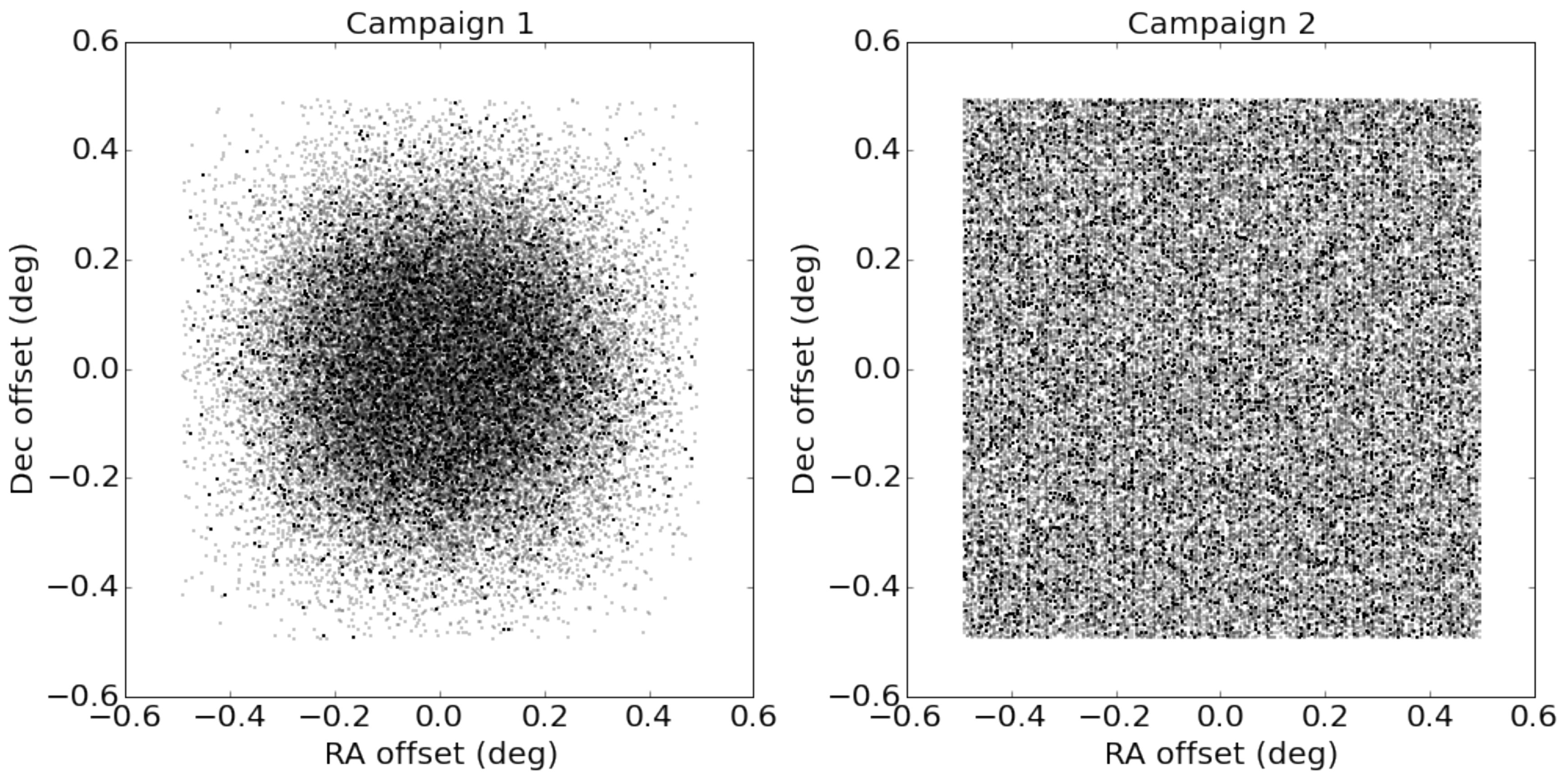}
\caption{Examples of the spatial distribution of candidates (all from thermal noise) that were detected in the first campaign (left panel) and second campaign (right panel). No primary beam correction has been applied in this analysis. Intead, the spatial structure is introduced by the spectral modulation filter used in the first campaign. Test observations of a pulsar show that this filter does not affect true candidates. \label{candlocall}}
\end{center}
\end{figure}

Images were not corrected for the varying primary beam gain in order to have uniform noise and equal chance of detection candidates throughout an image. However, as shown in Figure \ref{candlocall}, the spatial distribution of candidates in the first campaign is centrally concentrated. This effect is introduced by the rephasing of visibilities when measuring the spectral modulation of candidates. Rephasing is done with exact $uv$ coordinates as opposed to gridded coordinates used by FFT images and this introduces small random-like phase changes that cause fluctuations in the brightness of thermal-noise candidates. These phase changes scale with offset location, so thermal-noise candidates far from the phase center are affected more strongly. True candidates (with an underlying, coherent signal) are not affected. We have confirmed this by checking that the number of rejected candidates detected toward pulsar B0355+54 does not change with offset location. We also see that the significance of a true B0355+54 pulse does not change after rephasing to its location. Generally, we found that thermal-noise candidates were more sensitive to boundary effects (i.e., assumed DM, time, and \emph{uv} grid).

Visual inspection of the normal quantile plots typically found one candidate that exceeded thermal noise expectations for every few hours of data. These candidates had an S/N about 10\% higher than the S/N of the peak thermal-noise candidate; outside of known pulsar pulse, no candidates brighter than $10\sigma$ were ever seen. Manual inspection of candidates showed that most suffered from obvious RFI or edge effects (e.g., telescope slewing during data acquisition) and they have been removed from Figure \ref{normprob}. All other candidates were rejected because the measured S/N was very sensitive to flagging. We also tried to reproduce some candidates with a finer DM and \emph{uv} grid on the expectation that grid effects can depress sensitivity to candidates that fall at DM/pixel boundaries. In all cases, the candidate S/N dropped rapidly with small changes in DM or image gridding, indicating that the signal was not coherent across time, channel, and \emph{uv} space, and thus was likely generated by thermal noise.

\begin{figure}[htb]
\begin{center}
\includegraphics[width=0.98\columnwidth]{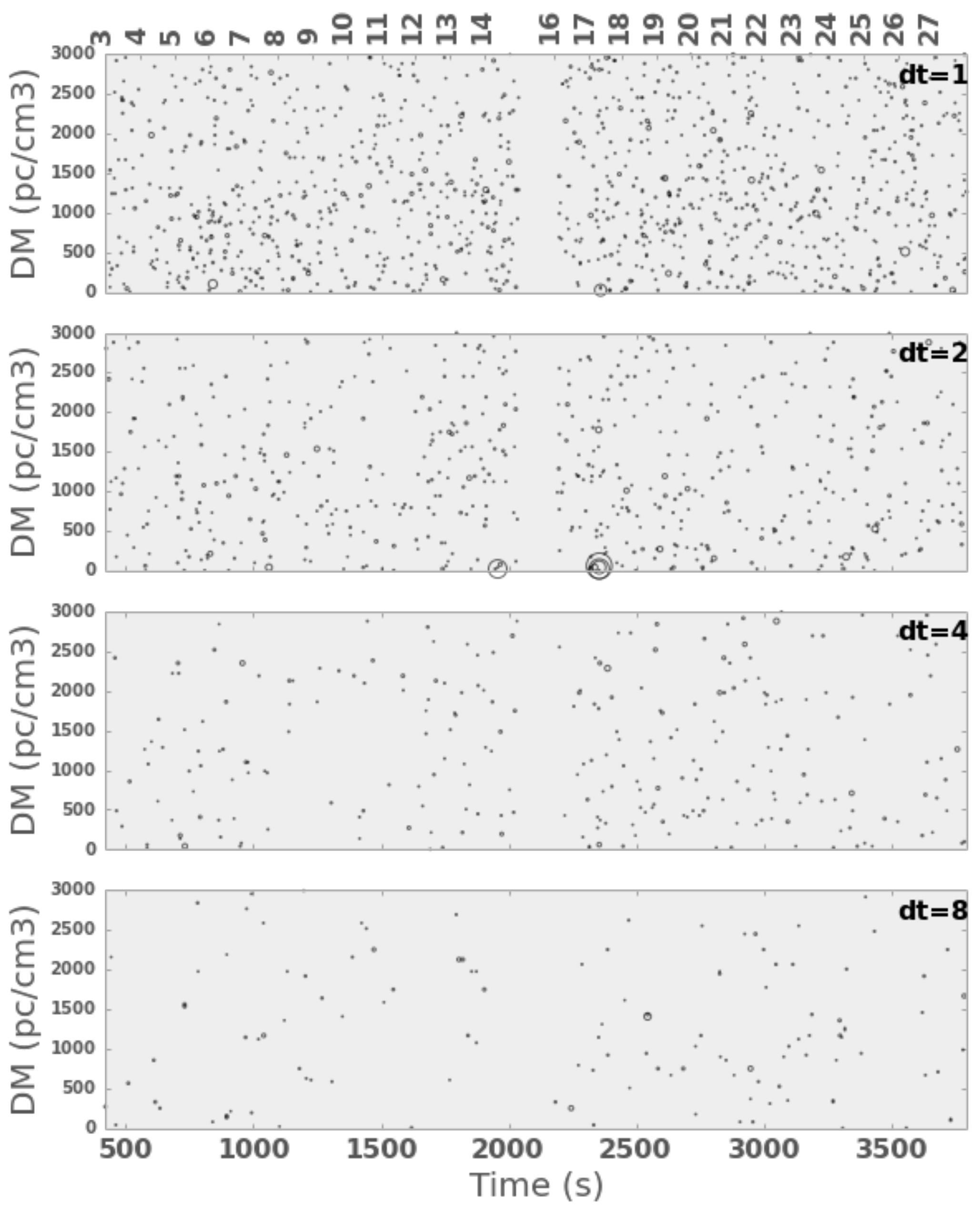}
\caption{DM versus time (bottom label in seconds from start, top label in scan number) for candidates from an observation on July 1, 2014. The four panels show the candidates at timescales corresponding to 1, 2, 4, and 8 integrations, or 5, 10, 20, and 40 ms widths. Symbol size scales with candidate S/N from values of 6.0 to 8.0. Images of the brightest events show that they were associated with a known Galactic puslar. Scan 15 observed a gain calibrator and was not searched for transients. \label{dmt}}
\end{center}
\end{figure}

Figure \ref{dmt} shows an example of how candidates are distributed as a function of DM and time. The plot shows the roughly 2000 candidates detected brighter than $6.0\sigma$ on four timescales during an hour-long observation in D configuration. In this case, the field was centered on a known Galactic pulsar, PSR J2248-0101, which has a DM of 29 pc cm$^{-3}$. Images for the few high-S/N, low-DM candidates seen in Figure \ref{dmt} show that they originated from that pulsar. Overall, about 20 significant transients were detected from the two faint pulsars in our target fields. In all cases, the transients were only seen at the center of the image with low DM ($<100$ pc cm$^{-3}$) and were ignored in all subsequent analysis. 

%

\section{Calculating Survey Effectiveness}
\label{qual}
\subsection{Completeness and Quality Checks}
Our search found no new transients in 166 hours of observing of high Galactic latitude fields. Here, we use data quality metrics to measure the survey completeness and place a constraint on the FRB rate.

We measured data quality at regular intervals throughout the search. Image quality was measured by the standard deviation of pixel values, which was typically clustered near the theoretical limit with a tail of higher values due to sporadic interference or bad calibration. Our flux-calibrated observations had a median image noise of roughly 13 mJy beam$^{-1}$, as expected for 5-ms images made with data from 26 good antennas, 232 MHz of bandwidth centered near 1.4 GHz. Since our candidate event rate is limited by thermal noise and that rate varies with image size, our threshold varied between field/configuration. The typical flux limit of an observation ranged from 7--8$\sigma$, or 90--100 mJy beam$^{-1}$ in 5 ms.

Formally, our completeness limit is worse than the naive estimate, since our DM grid allows up to 25\% sensitivity loss between DM values. However, our DM prescription is typical for this kind of transient search \citep[e.g.,][]{2013Sci...341...53T}, so relative rate comparisons will be unaffected by this loss. On the other hand, since our observations do not resolve the pulse in time, our search is not subject to sensitivity loss due to pulses falling on boundaries of the standard boxcar smoothing algorithm \citep[``dedisperse-all'';][]{2014arXiv1409.6125K}. As described in \S \ref{rate}, this makes the VLA sensitivity approximately equal to the naive flux limit of a comparable single-dish observation.

Gain calibration quality was checked by applying calibration solutions to calibrators and imaging with our custom imager. The calibration solutions and image quality were found to be stable in time. Bandpasses were found to be stable between observations, as expected.

Our pulsar test observations were flux calibrated and the faintest pulses detected were consistent with the separate data quality measurements. We also imaged a single pulse over a range of \emph{uv} gridding options to test whether pixel boundaries can reduce sensitivity. We found that the detected pulse S/N varied by $1\sigma$, consistent with noise-like variation.

\subsection{Measured Field of View}
Nominally, the field of view is equal to the ratio of observing wavelength to the diameter of an individual dish. The sensitivity pattern on the sky (``primary beam'') is approximated by a Gaussian. At our central observing frequency of 1.4 GHz, the VLA primary beam FWHM is 30\arcmin. Since the field of view changes with frequency, our effective field of view depends on the spectral index of the source. Imaging algorithms can also alter the effective field of view though choices of \emph{uv} gridding and corrections for wide fields of view \citep{1992A&A...261..353C}.

Rather than appealing to expected performance, we directly measured the end-to-end transient detection efficiency over the entire field of view. We did this with 8 sets of observation of the bright pulsar B0355+54. During each observation, we pointed at four positions offset in declination by 0\arcmin, 7\arcmin, 15\arcmin\ (the half power point), and 25\arcmin. By comparing the rate of pulse detection in each pointing, we measure the effect of primary beam sensitivity, \emph{uv} gridding, and candidate filtering.

To do this, we must assume that the pulse brightness distribution follows a powerlaw. We expect that the pulsar's \emph{intrinsic} flux density is stable on time scales much longer than our observations \citep{2000ApJ...539..300S}. However, scintillation will induce variations in the pulsar's intensity on faster timescales. Using the scintillation values compiled by \citet{1995ApJ...451..717G} and and formulae from \citet{1991ApJ...376..123C}, we estimate that diffractive scintillation will induce 10\%--30\% variation on timescales of roughly 2 minutes, or about 2 times longer than our pulsar scan duration. This is consistent with previous observations \citep{2006ChJAS...6b.204S}. On much longer time scales (weeks), we also expect an additional, roughly 10\% intensity variations due to refractive scintillations. Changes in the scintillation on this timescale will introduce discontinuities in the pulse brightness distribution, which we discuss below. 

Assuming pulse brightnesses are drawn from a single, powerlaw distribution, the rate of detection of pulses is:
\begin{equation}
R(F, d) \propto (F/\epsilon_d)^{-k},
\end{equation}
\noindent where $F$ is the flux limit, $\epsilon_d$ is the normalized primary beam gain (ranging from 0 to 1) for an offset pointing distance $d$, and $k$ is the powerlaw slope of the brightness distribution. By measuring the rate at a fixed flux limit, we can use the ratio of fluxes at different pointings to measure $\epsilon_d$.
\begin{equation}
F(R0, d) / F(R0, 0) = \epsilon_d.
\end{equation}

\begin{figure}[htb]
\begin{center}
\includegraphics[width=0.98\columnwidth]{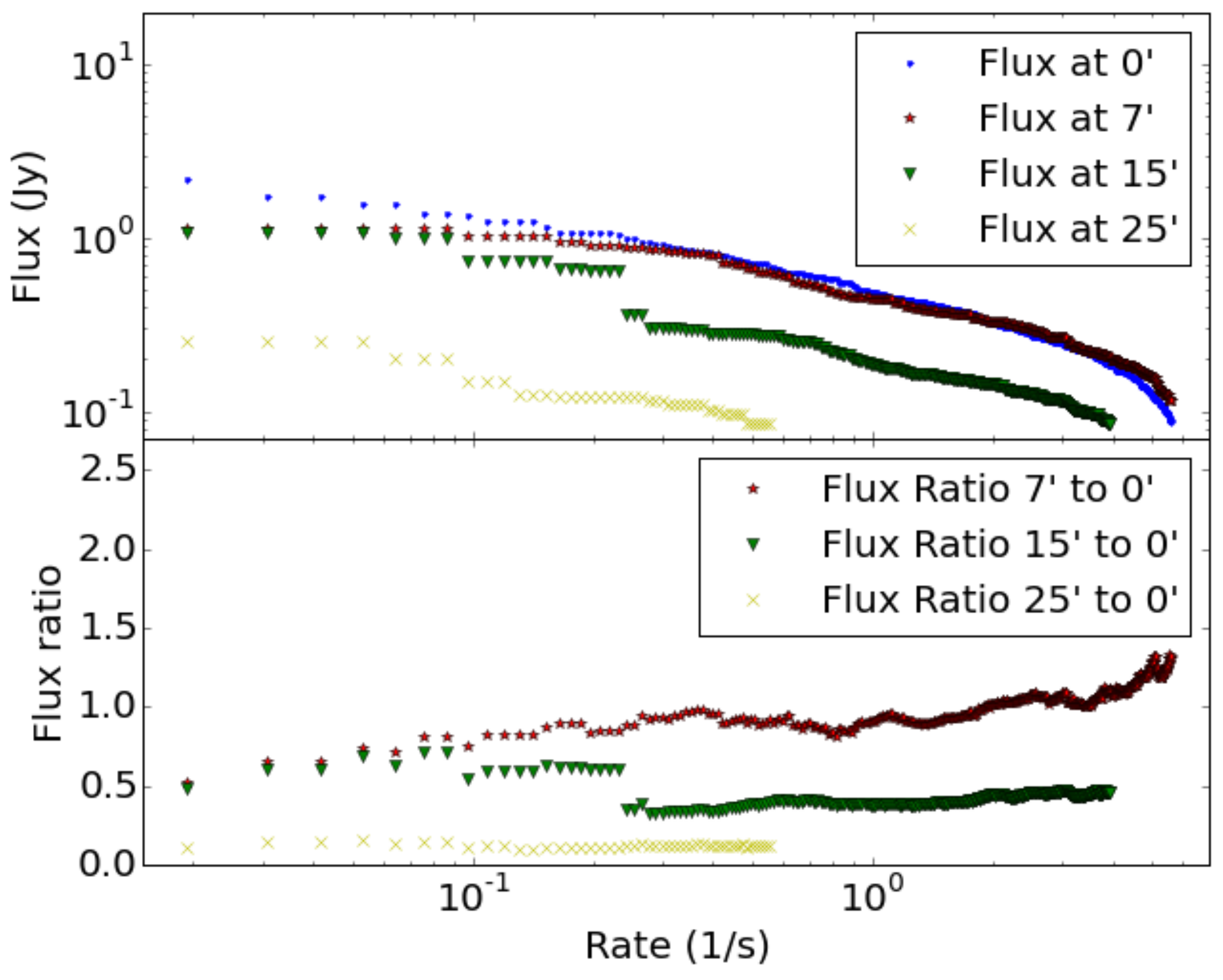}
\caption{(Top) Measured rate of detection of pulses from B0355+54 as a function of limiting sensitivity for an observation on 26 September 2013 (CnB configuration). Four colored lines each show the result of 1 minute of observing located at pointing center (blue), 7' off (red), 15' off (green), and 25' off (yellow). For a given rate, the pulsar is weaker in the offset pointing due to primary beam attenuation and other imaging artifacts. (Bottom) The ratio of the measured flux for the offset observations relative to that at the pointing center. \label{pulsarrate}}
\end{center}
\end{figure}

The top panel of Figure \ref{pulsarrate} shows the measured brightness of B0355+54 pulses as a function of pulse rate for a single set of scans. The brightest (rarest) part of the pulse brightness distribution is found toward the left of the plot and can be approximated by a powerlaw. The bottom panel shows the ratio of apparent pulse brightness of the offset pointings to the centered pointing, which we use to estimate $\epsilon_d$. We tested multiple methods of estimating the flux ratio and ultimately chose to measure it at a single rate corresponding to the faintest pulse detected in all pointings. The details of the method used had little effect on the measured value for $\epsilon_d$.

\begin{figure}[htb]
\begin{center}
\includegraphics[width=0.98\columnwidth]{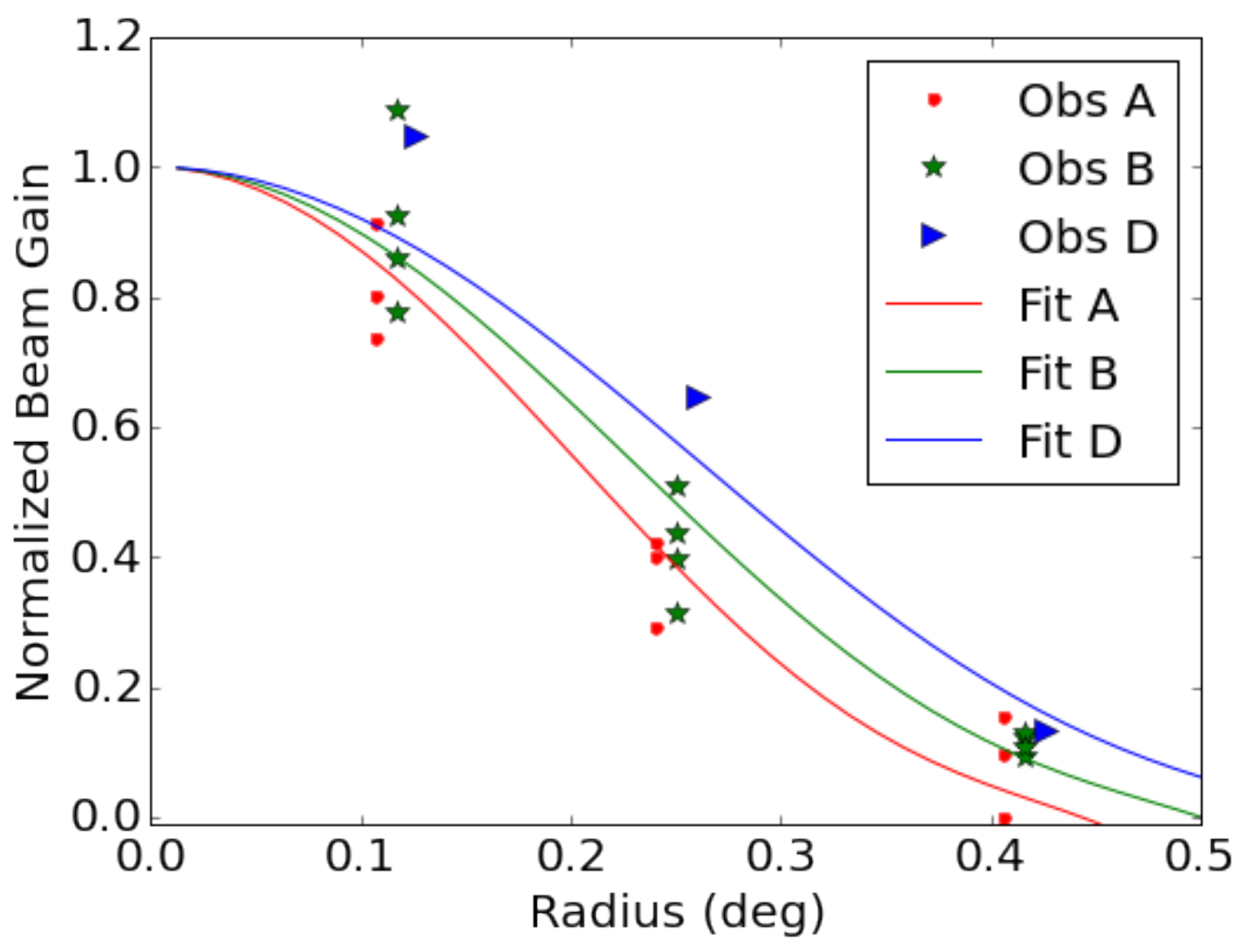}
\caption{Nominal and measured field of view of the VLA based on pulsar test observations. Stars show the measured values, while lines show a best-fit model (as defined by R. Perley and reported in AIPS PBCOR documentation). The three colors are used to represent data and model for each antenna configuration with pulsar test observations. \label{beamshape}}
\end{center}
\end{figure}

The mean value of $\epsilon_d$ for each configuration is shown in Table \ref{pulsartest}. This test showed that our system found pulses throughout the field of view, but that the effective field of view varies slightly with array configuration. This is consistent with the effect of the non-coplanar baselines on image quality for our simple (pillbox) gridding algorithm. Figure \ref{beamshape} visualizes the pulsar test observations and our estimate of the effective field of view for each antenna configuration. Repeated measurements of $\epsilon_d$\ in A and B configuration show that individual measurements are uncertain by $\pm0.1$. Given the stochastic nature of scintillation and its mode changes, it is difficult to formulate a more formal error analysis. 

\begin{table}
\caption{Pulsar Test Observations}
\centering
\begin{tabular}{cc|ccc}
Date              & Config & $\epsilon_{7\arcmin}$ & $\epsilon_{15\arcmin}$ & $\epsilon_{25\arcmin}$ \\ \hline
25 Sep 2013  & B & \multirow{4}{*}{0.91} & \multirow{4}{*}{0.41} & \multirow{4}{*}{0.11} \\
26 Sep 2013  & B & & & \\
13 Dec 2013  & B & & & \\ 
11 Jan 2014   & B & & & \\ \hline
16 Jun 2014   & A & \multirow{3}{*}{0.82} & \multirow{3}{*}{0.37} & \multirow{3}{*}{0.08} \\
17 Jun 2014   & A & & & \\
21 Jun 2014   & A & & & \\ \hline
3 Sep 2014    & D & 1.05 & 0.65 & 0.13 \\ \hline
\multicolumn{2}{c|}{VLA beam (band center)} & 0.88 & 0.53 & 0.13 \\
\end{tabular}
\label{pulsartest}
\end{table}

We fit a VLA primary beam model to the pulsar $\epsilon_d$ measurements to estimate the effective field of view in each configuration. We do not have a model that incorporates all the effects that may reduce the field of view, so we treat the observing frequency of the primary beam model as a free parameter. The best-fit model in A, B, and D configurations has a field of view with an effective observing frequency of 1.67, 1.48, and 1.3 GHz, respectively; these models have FWHM of 13.0\arcmin, 14.7\arcmin, and 16.8\arcmin, respectively.

In D configuration, we expect no effect from non-coplanar baselines, so the effective observing frequency should be the true center frequency of 1.396 GHz. The effective field of view in D configuration is larger than that at band center, reflecting the fact that B0355+54 has a spectral index of --0.9 \citep{1995MNRAS.273..411L}. Thus, this estimate of the survey speed effectively assumes a source with a spectral index of --0.9. This is discussed further in \S \ref{caveats}.

\section{VLA Limit on FRB Rate}
\label{rate}
\subsection{Data Quality and Survey Speed}
Using the VLA survey, we place a limit on the rate of FRBs above any specified fluence value. This section describes how we deal with (1) different flux sensitivity limits across observations, (2) variations in S/N significance limits across observations, and (3) dependence of the field of view area on its gain limit and on the array configuration.

For each configuration, the observations with flux calibrators are used to produce a set of Jy-scaled images. For each image, the standard deviation of background pixels is calculated ($F_{ij}$\ for image $j$\ in observation $i$); these represent the distribution of sensitivity (Jy) at the $1\sigma$\ level across all observations in that configuration. In addition, the flux limits are scaled by the highest thermal-noise S/N value ($N_i$), which is measured in each observation. These values are the largest S/N values shown in Figure \ref{normprob}, typically ranging from 7 to 8.

\begin{figure}[htb]
\begin{center}
\includegraphics[width=0.98\columnwidth]{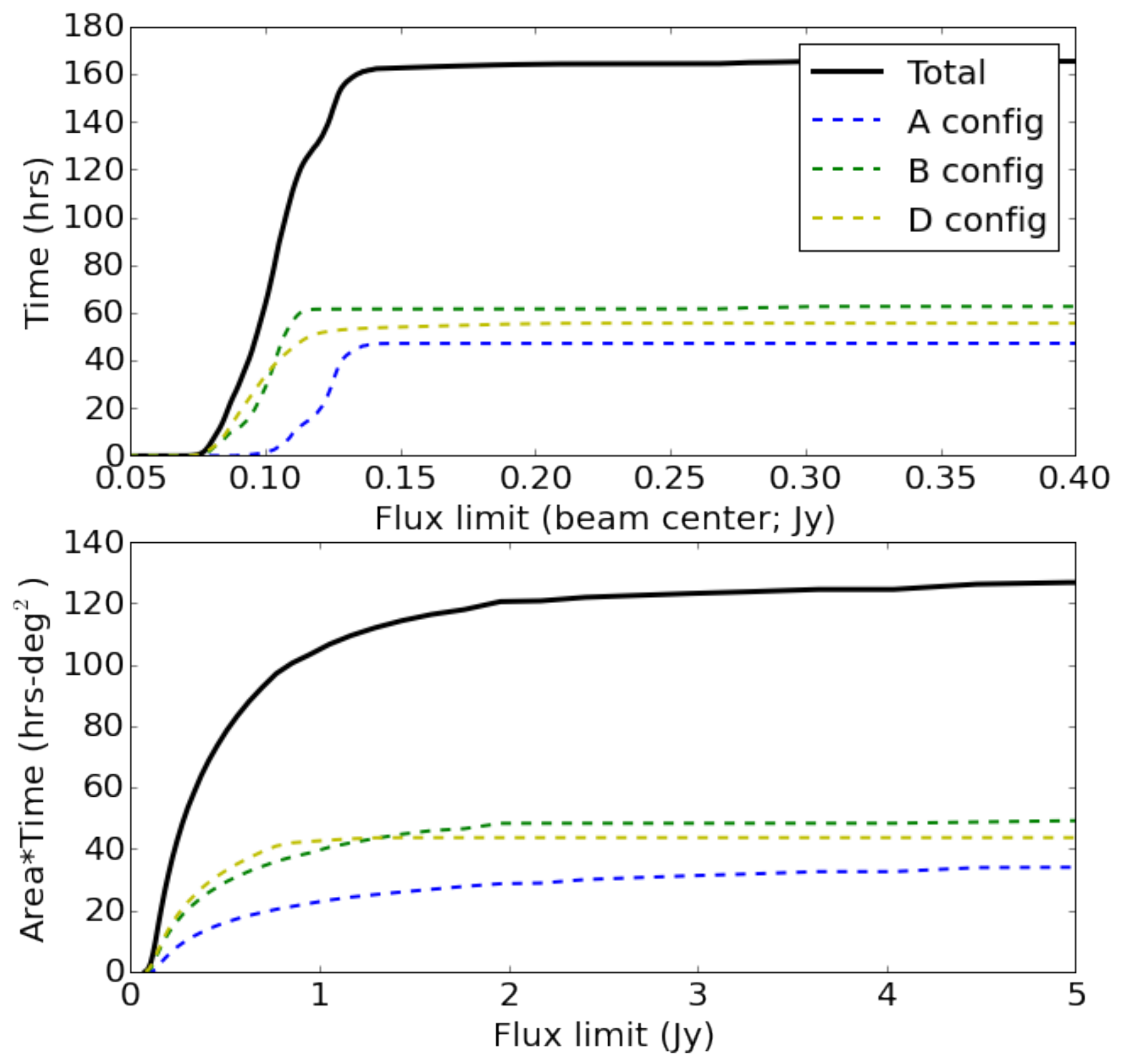}
\caption{(Top) Cumulative histogram of observing time as a function of flux limit. No primary beam correction has been applied, so these values are strictly only correct at the center of the primary beam. The dashed lines show the histogram for observations in A, B, and D configuration and the solid line shows their sum. (Right) The maximum of the product of primary beam area and observing time as a function of flux limit. \label{fluxlims}}
\end{center}
\end{figure}

The time on sky with complete detection sensitivity at least F (Jy; beam center) is estimated with a cumulative sum as
\begin{equation}
T_c(F) = H_c \cdot n_c^{-1} \sum_{i,j} I(F_{c,i,j}N_{c,i} < F_c)
\end{equation}
where $I$\ is the indicator function equal to 1 if its argument is true and 0 otherwise; $H_c$\ is the hours in configuration $c$ (grouped for observations closest to A, B, and D configurations), and $n$\ is the total number of calibrator images. These cumulative time curves and their sum are shown in the top panel of Figure \ref{fluxlims}. Given that observations in A configuration used two-stage imaging, the S/N threshold changed as more antennas were included in the first-stage image. In this case, the less sensitive of the two thresholds was used to define the limit, $N_i$.

The area in the field of view depends on the gain limit used in its definition. It is typically assumed that the primary beam FWHM ($\epsilon=0.5$) defines the survey field of view, but that understates the potential for discovery in the small, highly-sensitive area near the center of the primary beam and the large, low-sensitivity areas far out in the primary beam. To cover all these cases, we first calculate the area, $A_c(\epsilon)$ (in degrees$^2$), with normalized gain greater than $\epsilon$. For example, Figure \ref{beamshape} shows for configuration D, $A_D(0.5)=\pi \cdot 0.27^2$\ degrees$^2$. The area-time product with sensitivity to all sources of flux $F$\ or larger at a relative gain of $\epsilon$\ is $A_c(\epsilon)T_c(\epsilon F)$. We consider the full range of primary beam sensitivity and data quality by finding the maximum area-time product ($\Omega$) for a given limiting sensitivity $F$ as
\begin{equation}
\Omega_c(F) = \max_{\epsilon \in (0,1]} A_c (\epsilon) T_c (\epsilon F).
\end{equation}
These curves are shown in the bottom panel of Figure \ref{fluxlims}.

Finally, the estimated rate of FRBs (events per time per area) above a flux limit $F$\ is
\begin{equation}
R(F) = \frac{N}{\Omega_A(F) + \Omega_B(F) + \Omega_D(F)}
\label{rateeqn}
\end{equation}

\noindent where $N$ is the number of detections. With zero detections, Poisson-based upper 100$\alpha$\% confidence limits on the rate are obtained by taking $N=-\log(1-\alpha)$. For example, 95\% limits use $N=-\log(0.05)=3.0$ and 50\% limits use $N=-\log(0.5)=0.693$. Figure \ref{rateplot} shows these confidence limit curves in red.

\subsection{VLA rate limit vs.\ Published rates}
\begin{figure*}[htb]
\begin{center}
\includegraphics[width=2\columnwidth]{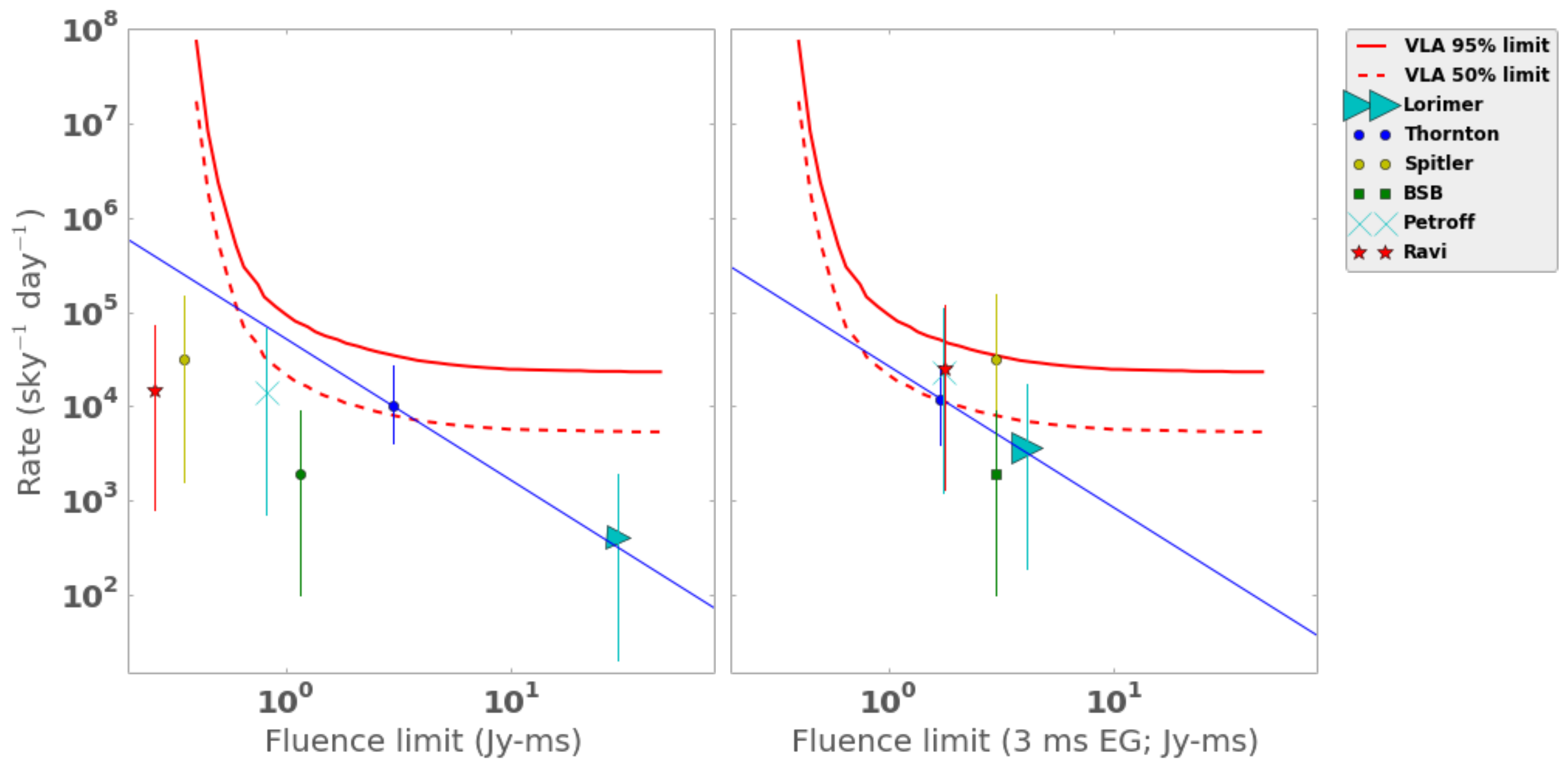}
\caption{ \emph{Left:} FRB rates and VLA rate limit as a function of limiting fluence as quoted in (or inferred from) publications \citep{2007Sci...318..777L,2013Sci...341...53T, 2014ApJ...790..101S, 2014ApJ...792...19B,2014arXiv1412.1599R,2014arXiv1412.0342P}. The blue line shows an extrapolation of the rate of \citet{2013Sci...341...53T}, assuming a Euclidean distribution ($-3/2$\ powerlaw slope in this space) from a fluence limit of 3 Jy-ms. The VLA 50\% and 95\% upper limits are shown with a dashed and solid line, respectively. The VLA rate limit is not complete at very high fluences (far right) as described in \S \ref{caveats}. The fluence limit of \citet{2007Sci...318..777L} is a lower limit recently calculated in a reanalysis \citep{2014arXiv1409.6125K}. \emph{Right:} Same as the left panel, but for recalculated flux limits. All flux limits assume a 3 ms pulse width and that FRBs originate outside the Galaxy with DM of 779 pc cm$^{-3}$. \label{rateplot}}
\end{center}
\end{figure*}

The left panel of Figure \ref{rateplot} compares the VLA rate limit as a function of fluence limit to FRB rates quoted in publications. In constructing this figure, we found that the sensitivity of most published surveys are defined heterogeneously. \citet{2014ApJ...790..101S} calculate a flux limit using the mean beam gain within the FWHM\footnote{For this FRB, we consider the rate implied by the area covered by the main beam and sidelobes, which is most consistent with the unusual spectral index of the detection.}. \citet{2007Sci...318..777L} use the measured fluence of their detection to define a fluence limit. Finally, \citet{2013Sci...341...53T} don't report a fluence limit at all, but instead measure the mean fluence of all detections. In general, the sensitivity should be defined at the completeness limit, which is 2 times the ideal sensitivity for an area calculated within the FWHM.

Assuming a Euclidean extrapolation on the rate of \citet{2013Sci...341...53T}, the VLA rate limit is most constraining for an area defined by the primary beam FWHM. The nominal image sensitivity is 120 mJy in 5 ms and the limiting sensitivity in the FWHM is 240 mJy in 5 ms or 1.2 Jy-ms. At that limit, the VLA 95\% confidence upper limit on the FRB rate is $7\times10^4$ sky$^{-1}$ day$^{-1}$, equivalent to a $\sim81$\% constraint on the published FRB rate. This limit is strengthened when considering that the fluence limit of \citet{2013Sci...341...53T} assumed ideal sensitivity and did not correct for known sensitivity losses due to the dedispersion algorithm used \citep[``dedisperse-all'';][]{2014arXiv1409.6125K}. Correcting for these effects is difficult, but they suggest that the VLA rate constraint is not consistent with the published rate of \citet{2013Sci...341...53T}.

\subsection{A Homogenous FRB Rate Definition}
\label{recalculated}
However, the heterogeneity of the methods used to calculate the published fluence limits makes it difficult to compare these rates. Time and spectral resolution affect the sensitivity as a function of DM and scattering. This sensitivity varies as a function of Galactic latitude, which has been used to explain the observed deficit of FRB detections at low and intermediate Galactic latitudes \citep{2014ApJ...792...19B,2014arXiv1405.5945P}. To enable a more meaningful comparison of FRB rates, we introduce two techniques to standardize the calculation of flux and rate. First, we define the flux limit at the primary beam half-power point, which is the sensitivity to which a survey covered by the FWHM is complete. Second, we calculate the sensitivity of each survey in units of Jy and incorporate factors that can reduce detection efficiency (e.g., sky noise, dispersion smearing) via the radiometer equation. This approach is described in \citet{2014ApJ...792...19B}, which calculates the relative number of FRBs detected in past surveys. We extend this technique to the calculation of an effective noise for an entire survey, which allows us to scale previous detections from units of S/N to flux in Jy. The assumption of a single effective noise for a survey is valid when the survey was conducted over regions with roughly constant properties (sky temperature, DM, etc.), which is true for the low-dispersion and high-Galactic-latitude surveys we discuss here.

The right panel of Figure \ref{rateplot} uses this new approach to compare the VLA rate limit to previous FRB rate measurements. The sensitivity calculation assumes that pulses have DM and widths equal to the mean of several typical FRBs \citep[779 pc cm$^{-3}$ and 3 ms;][]{2013Sci...341...53T, 2014ApJ...792...19B, 2014ApJ...790..101S}. The new flux limit estimates show that published rates are more consistent with single population with a Euclidean flux distribution. This rate is roughly $1.2\times10^{4}$ sky$^{-1}$ day$^{-1}$ at a fluence limit of 1.7 Jy-ms. This effectively reduces the rate of \citet{2013Sci...341...53T}, which defined its rate for a mean fluence of 3 Jy-ms. 

\citet{2014arXiv1409.6125K} argue for a similar, reduced rate noting that estimates from past surveys did not consider the effect of pulse width on FRB detectability. We opt here for a simpler approach of calculating a rate for a single prototypical FRB. Unlike past single-dish data, the VLA data likely do not resolve the pulses in time and its sensitivity scales differently with pulse width. The newest Parkes flux limits are roughly equal to our VLA flux limit for a pulse width of 1 ms, while the VLA limit is 2 times lower (better) for a width of 5 ms.

Given our reduced FRB rate estimate, the VLA observations are less constraining than initially believed. The VLA campaign had a roughly 60\% chance of detecting an FRB, for an assumed width of 3 ms. Despite this downward revision in the apparent rate, the rate per galaxy defined in \citet{2013Sci...341...53T} is unchanged, since it was defined in terms of a DM-inferred distance.

\subsection{Caveats}
\label{caveats}
Our VLA rate constraint differs from that of previous work in that it includes measured fluxes, estimates of data quality, and end-to-end measurements of the field of view. Single-dish rate estimates typically assume ideal, thermal-noise limited data at all times. This likely biases their rate estimate below the true rate. In this regard, the VLA rate constraint is likely stronger than the formal estimate presented here, perhaps by a factor as high as $\sqrt{2}$ or roughly 40\%.

On the other hand, the field of view measurement introduces an assumed spectral index for FRBs. The field of view was inferred from the pulses detected toward pulsar B0355+54, which has a spectral index of $-0.9$. While there are no reliable measurements of FRB spectral index, there are some indirect indications that they have flat or even positive spectral indices \citep{2014ApJ...790..101S}. If so, then the effective VLA field of view will be smaller than implied by our pulsar tests. Assuming a flat spectral index reduces the field of view by 7\% and the survey speed by 13\%. This increases our 95\% confidence constraint on the FRB rate to roughly $8\times10^4$ above a fluence of 1.2 Jy-ms. This limit would have a roughly 50\% constraint on the recalculated FRB rate, as described in \S \ref{recalculated}.

This FRB rate analysis implicitly assumes that FRBs have a pulse width of 5 ms or less. The transient search pipeline used in the first campaign lost sensitivity for transient longer than 5 ms. Dispersion smearing affects DMs between 1850 and 3000 pc cm$^{-3}$, for which we expect roughly 25\% loss of sensitivity. The distribution of observed FRB pulse widths and their DMs suggest these are reasonable assumptions.

Finally, we found two effects that may reduce sensitivity for pulses brighter than $50\sigma$. First, our dynamic flagging algorithm can flag very bright ($\gtrsim70\sigma$), off-axis sources, since they have a large visibility variance between baselines (see \S \ref{proc}). Second, the spectral modulation filter can reject candidates brighter than $50\sigma$, if the pulse occupies only part of the spectrum. For these reasons, we consider our detection efficiency valid for $\rm{S/N}<50$. The actual S/N limit may be much higher, but we can only test the algorithm with known pulsar pulses, the brightest of which have S/N$\sim50$.

\subsection{A Limit on FRB Recurrence}
Our survey observed the field coincident with FRB120127 for 32 hours, which allows us to constrain its recurrence rate. The FRB location precision is 14\arcmin, which is half of VLA primary beam FWHM. Within this area, the VLA has a limiting sensitivity of 170 mJy beam$^{-1}$ in 5 ms or roughly 40\% higher than the minimum fluence of FRB 120127. While the true FRB fluence is not known, assuming a mean gain correction would make it equal to the VLA fluence limit. If so, then the VLA limits its recurrence rate to less than 3.2 day$^{-1}$ at 95\% confidence. This is a 2.6 times lower recurrence rate than the least frequent RRAT \citep[e.g.,][]{2011MNRAS.415.3065K} and is comparable to the limit made in 50 hours of follow-up observing of FRB010724 \citep{2007Sci...318..777L}.

\section{Conclusions}
\label{conc}
We presented the results of a 166-hour survey for FRBs with a new millisecond imaging mode of the VLA. This new mode required the extensive development of the VLA correlator and building custom, parallelized software to search the 1 TB hour$^{-1}$ data stream using two compute clusters. A suite of pulsar test observations provided an end-to-end test of the detection efficiency of the transient search pipeline. These tests show that we were sensitive to FRBs with brightnesses of 120 mJy beam$^{-1}$ (at beam center) in 5 ms over a field of view with FWHM of roughly 0.5\arcdeg.

With no detection, the VLA data limit the FRB rate of occurrence to less than $7\times10^4$ sky$^{-1}$ day$^{-1}$ (95\% confidence) above a fluence limit of 1.2 Jy-ms. If we assume that the FRB flux distribution is Euclidean (i.e., a spatially uniform population), the VLA limit is likely not consistent with that of \citet{2013Sci...341...53T}. We reanalyze the FRB rate estimates of past surveys with detections to include a homogeneous definition of flux limit that includes pulse width, sky temperature, dispersion, and primary beam attenuation. This reanalysis predicts that the VLA observations had a roughly 60\% chance of detecting a typical FRB.

The marginal VLA limit on the revised FRB rate does not allow us to make a definitive statement about their nature. To exclude this rate at 95\% confidence would require roughly 3 times as much time on sky ($\sim500$ hours total). A stronger VLA limit on the FRB rate would support the hypothesis of \citet{2014ApJ...797...70K} that perytons and FRBs are a single phenomenon originating in the Earth's ionosphere\footnote{The detection with the Arecibo Observatory \citep{2014ApJ...790..101S} defines a minimum elevation of 430 km for FRBs.}. This idea could also be tested with our VLA data by searching auto-correlations for peryton-like transients. 

This campaign shows that interferometric imaging is a viable means to search for and localize FRBs. However, data management and the computational load of this survey are substantial. In the future, real-time transient detection could alleviate the challenges of observing at these extreme data rates. Detecting transients at the telescope in real time would enable triggered data recording and other forms of ``data triage'' to control the data flow, while maintaining sensitivity to rare, transient phenomena. Combining such a system with automated candidate data quality checks (e.g., via machine learning) would reduce the need for human review and allow rapid response to radio transients. SKA precursor telescopes will be powerful transient survey machines \citep{2009arXiv0910.2935B,2008ExA....22..151J}, but will need to manage the data deluge to access science at data rates above 1 TB hour$^{-1}$.

\section*{Acknowledgements}
We thank the VLA staff, particularly Martin Pokorny, Ken Sowinski, Vivek Dhawan, James Robnett, and Joan Wrobel, for working tirelessly to support this challenging observing mode. Peter Williams contributed with wide-ranging Python expertise. This project was supported by the University of California Office of the President under Lab Fees Research Program Award 237863. The National Radio Astronomy Observatory is a facility of the National Science Foundation operated under cooperative agreement by Associated Universities, Inc. Part of this research was carried out at the Jet Propulsion Laboratory, California Institute of Technology, under a contract with the National Aeronautics and Space Administration. This research used resources of the National Energy Research Scientific Computing Center, a DOE Office of Science User Facility supported by the Office of Science of the U.S. Department of Energy under Contract No. DE-AC02-05CH11231.

\bibliography{fasttrants.bib}

\begin{thebibliography}{41}
\expandafter\ifx\csname natexlab\endcsname\relax\def\natexlab#1{#1}\fi

\bibitem[{{Bagchi} {et~al.}(2012){Bagchi}, {Nieves}, \&
  {McLaughlin}}]{2012MNRAS.425.2501B}
{Bagchi}, M., {Nieves}, A.~C., \& {McLaughlin}, M. 2012, \mnras, 425, 2501

\bibitem[{{Bannister} {et~al.}(2012){Bannister}, {Murphy}, {Gaensler}, \&
  {Reynolds}}]{2012ApJ...757...38B}
{Bannister}, K.~W., {Murphy}, T., {Gaensler}, B.~M., \& {Reynolds}, J.~E. 2012,
  \apj, 757, 38

\bibitem[{{Booth} {et~al.}(2009){Booth}, {de Blok}, {Jonas}, \&
  {Fanaroff}}]{2009arXiv0910.2935B}
{Booth}, R.~S., {de Blok}, W.~J.~G., {Jonas}, J.~L., \& {Fanaroff}, B. 2009,
  ArXiv e-prints

\bibitem[{{Bregman}(2007)}]{2007ARA&A..45..221B}
{Bregman}, J.~N. 2007, \araa, 45, 221

\bibitem[{{Burke-Spolaor} {et~al.}(2011){Burke-Spolaor}, {Bailes}, {Ekers},
  {Macquart}, \& {Crawford}}]{2011ApJ...727...18B}
{Burke-Spolaor}, S., {Bailes}, M., {Ekers}, R., {Macquart}, J.-P., \&
  {Crawford}, III, F. 2011, \apj, 727, 18

\bibitem[{{Burke-Spolaor} \& {Bannister}(2014)}]{2014ApJ...792...19B}
{Burke-Spolaor}, S., \& {Bannister}, K.~W. 2014, \apj, 792, 19

\bibitem[{Chambers(1983)}]{chambers1983graphical}
Chambers, J. 1983, Graphical methods for data analysis, Chapman \& Hall
  statistics series (Wadsworth International Group)

\bibitem[{{Cordes} \& {Lazio}(1991)}]{1991ApJ...376..123C}
{Cordes}, J.~M., \& {Lazio}, T.~J. 1991, \apj, 376, 123

\bibitem[{{Cordes} \& {Lazio}(2002)}]{2002astro.ph..7156C}
{Cordes}, J.~M., \& {Lazio}, T.~J.~W. 2002, ArXiv Astrophysics e-prints

\bibitem[{{Cornwell} \& {Perley}(1992)}]{1992A&A...261..353C}
{Cornwell}, T.~J., \& {Perley}, R.~A. 1992, \aap, 261, 353

\bibitem[{{Deng} \& {Zhang}(2014)}]{2014ApJ...783L..35D}
{Deng}, W., \& {Zhang}, B. 2014, \apjl, 783, L35

\bibitem[{{Falcke} \& {Rezzolla}(2014)}]{2014A&A...562A.137F}
{Falcke}, H., \& {Rezzolla}, L. 2014, \aap, 562, A137

\bibitem[{{Fang} {et~al.}(2013){Fang}, {Bullock}, \&
  {Boylan-Kolchin}}]{2013ApJ...762...20F}
{Fang}, T., {Bullock}, J., \& {Boylan-Kolchin}, M. 2013, \apj, 762, 20

\bibitem[{{Gupta}(1995)}]{1995ApJ...451..717G}
{Gupta}, Y. 1995, \apj, 451, 717

\bibitem[{{Johnston} {et~al.}(2008){Johnston}, {Taylor}, {Bailes}, {Bartel},
  {Baugh}, {Bietenholz}, {Blake}, {Braun}, {Brown}, {Chatterjee}, {Darling},
  {Deller}, {Dodson}, {Edwards}, {Ekers}, {Ellingsen}, {Feain}, {Gaensler},
  {Haverkorn}, {Hobbs}, {Hopkins}, {Jackson}, {James}, {Joncas}, {Kaspi},
  {Kilborn}, {Koribalski}, {Kothes}, {Landecker}, {Lenc}, {Lovell}, {Macquart},
  {Manchester}, {Matthews}, {McClure-Griffiths}, {Norris}, {Pen}, {Phillips},
  {Power}, {Protheroe}, {Sadler}, {Schmidt}, {Stairs}, {Staveley-Smith},
  {Stil}, {Tingay}, {Tzioumis}, {Walker}, {Wall}, \&
  {Wolleben}}]{2008ExA....22..151J}
{Johnston}, S., {et~al.} 2008, Experimental Astronomy, 22, 151

\bibitem[{{Kashiyama} {et~al.}(2013){Kashiyama}, {Ioka}, \&
  {M{\'e}sz{\'a}ros}}]{2013ApJ...776L..39K}
{Kashiyama}, K., {Ioka}, K., \& {M{\'e}sz{\'a}ros}, P. 2013, \apjl, 776, L39

\bibitem[{{Keane} {et~al.}(2011){Keane}, {Kramer}, {Lyne}, {Stappers}, \&
  {McLaughlin}}]{2011MNRAS.415.3065K}
{Keane}, E.~F., {Kramer}, M., {Lyne}, A.~G., {Stappers}, B.~W., \&
  {McLaughlin}, M.~A. 2011, \mnras, 415, 3065

\bibitem[{{Keane} \& {Petroff}(2014)}]{2014arXiv1409.6125K}
{Keane}, E.~F., \& {Petroff}, E. 2014, ArXiv e-prints

\bibitem[{{Keith} {et~al.}(2010){Keith}, {Jameson}, {van Straten}, {Bailes},
  {Johnston}, {Kramer}, {Possenti}, {Bates}, {Bhat}, {Burgay}, {Burke-Spolaor},
  {D'Amico}, {Levin}, {McMahon}, {Milia}, \& {Stappers}}]{2010MNRAS.409..619K}
{Keith}, M.~J., {et~al.} 2010, \mnras, 409, 619

\bibitem[{{Kocz} {et~al.}(2012){Kocz}, {Bailes}, {Barnes}, {Burke-Spolaor}, \&
  {Levin}}]{2012MNRAS.420..271K}
{Kocz}, J., {Bailes}, M., {Barnes}, D., {Burke-Spolaor}, S., \& {Levin}, L.
  2012, \mnras, 420, 271

\bibitem[{{Kulkarni} {et~al.}(2014){Kulkarni}, {Ofek}, {Neill}, {Zheng}, \&
  {Juric}}]{2014ApJ...797...70K}
{Kulkarni}, S.~R., {Ofek}, E.~O., {Neill}, J.~D., {Zheng}, Z., \& {Juric}, M.
  2014, \apj, 797, 70

\bibitem[{{Law} \& {Bower}(2012)}]{2012ApJ...749..143L}
{Law}, C.~J., \& {Bower}, G.~C. 2012, \apj, 749, 143

\bibitem[{{Law} {et~al.}(2012){Law}, {Bower}, {Pokorny}, {Rupen}, \&
  {Sowinski}}]{2012ApJ...760..124L}
{Law}, C.~J., {Bower}, G.~C., {Pokorny}, M., {Rupen}, M.~P., \& {Sowinski}, K.
  2012, \apj, 760, 124

\bibitem[{{Lorimer} {et~al.}(2007){Lorimer}, {Bailes}, {McLaughlin},
  {Narkevic}, \& {Crawford}}]{2007Sci...318..777L}
{Lorimer}, D.~R., {Bailes}, M., {McLaughlin}, M.~A., {Narkevic}, D.~J., \&
  {Crawford}, F. 2007, Science, 318, 777

\bibitem[{{Lorimer} {et~al.}(1995){Lorimer}, {Yates}, {Lyne}, \&
  {Gould}}]{1995MNRAS.273..411L}
{Lorimer}, D.~R., {Yates}, J.~A., {Lyne}, A.~G., \& {Gould}, D.~M. 1995,
  \mnras, 273, 411

\bibitem[{{Macquart} \& {Koay}(2013)}]{2013ApJ...776..125M}
{Macquart}, J.-P., \& {Koay}, J.~Y. 2013, \apj, 776, 125

\bibitem[{{McLaughlin} \& {Cordes}(2003)}]{2003ApJ...596..982M}
{McLaughlin}, M.~A., \& {Cordes}, J.~M. 2003, \apj, 596, 982

\bibitem[{{McMullin} {et~al.}(2007){McMullin}, {Waters}, {Schiebel}, {Young},
  \& {Golap}}]{2007ASPC..376..127M}
{McMullin}, J.~P., {Waters}, B., {Schiebel}, D., {Young}, W., \& {Golap}, K.
  2007, in Astronomical Society of the Pacific Conference Series, Vol. 376,
  Astronomical Data Analysis Software and Systems XVI, ed. R.~A. {Shaw},
  F.~{Hill}, \& D.~J. {Bell}, 127

\bibitem[{{Metzger} {et~al.}(1997){Metzger}, {Djorgovski}, {Kulkarni},
  {Steidel}, {Adelberger}, {Frail}, {Costa}, \&
  {Frontera}}]{1997Natur.387..878M}
{Metzger}, M.~R., {Djorgovski}, S.~G., {Kulkarni}, S.~R., {Steidel}, C.~C.,
  {Adelberger}, K.~L., {Frail}, D.~A., {Costa}, E., \& {Frontera}, F. 1997,
  \nat, 387, 878

\bibitem[{{Palaniswamy} {et~al.}(2014){Palaniswamy}, {Wayth}, {Trott},
  {McCallum}, {Tingay}, \& {Reynolds}}]{2014ApJ...790...63P}
{Palaniswamy}, D., {Wayth}, R.~B., {Trott}, C.~M., {McCallum}, J.~N., {Tingay},
  S.~J., \& {Reynolds}, C. 2014, \apj, 790, 63

\bibitem[{{Petroff} {et~al.}(2014{\natexlab{a}}){Petroff}, {Bailes}, {Barr},
  {Barsdell}, {Bhat}, {Bian}, {Burke-Spolaor}, {Caleb}, {Champion}, {Chandra},
  {Da Costa}, {Delvaux}, {Flynn}, {Gehrels}, {Greiner}, {Jameson}, {Johnston},
  {Kasliwal}, {Keane}, {Keller}, {Kocz}, {Kramer}, {Leloudas}, {Malesani},
  {Mulchaey}, {Ng}, {Ofek}, {Perley}, {Possenti}, {Schmidt}, {Shen},
  {Stappers}, {Tisserand}, {van Straten}, \& {Wolf}}]{2014arXiv1412.0342P}
{Petroff}, E., {et~al.} 2014{\natexlab{a}}, ArXiv e-prints

\bibitem[{{Petroff} {et~al.}(2014{\natexlab{b}}){Petroff}, {van Straten},
  {Johnston}, {Bailes}, {Barr}, {Bates}, {Bhat}, {Burgay}, {Burke-Spolaor},
  {Champion}, {Coster}, {Flynn}, {Keane}, {Keith}, {Kramer}, {Levin}, {Ng},
  {Possenti}, {Stappers}, {Tiburzi}, \& {Thornton}}]{2014arXiv1405.5945P}
---. 2014{\natexlab{b}}, ArXiv e-prints

\bibitem[{{Ravi} {et~al.}(2014){Ravi}, {Shannon}, \&
  {Jameson}}]{2014arXiv1412.1599R}
{Ravi}, V., {Shannon}, R.~M., \& {Jameson}, A. 2014, ArXiv e-prints

\bibitem[{{Rees}(1977)}]{1977Natur.266..333R}
{Rees}, M.~J. 1977, \nat, 266, 333

\bibitem[{{Saint-Hilaire} {et~al.}(2014){Saint-Hilaire}, {Benz}, \&
  {Monstein}}]{2014ApJ...795...19S}
{Saint-Hilaire}, P., {Benz}, A.~O., \& {Monstein}, C. 2014, \apj, 795, 19

\bibitem[{{Siemion} {et~al.}(2012){Siemion}, {Bower}, {Foster}, {McMahon},
  {Wagner}, {Werthimer}, {Backer}, {Cordes}, \& {van
  Leeuwen}}]{2012ApJ...744..109S}
{Siemion}, A.~P.~V., {et~al.} 2012, \apj, 744, 109

\bibitem[{{Spitler} {et~al.}(2012){Spitler}, {Cordes}, {Chatterjee}, \&
  {Stone}}]{2012ApJ...748...73S}
{Spitler}, L.~G., {Cordes}, J.~M., {Chatterjee}, S., \& {Stone}, J. 2012, \apj,
  748, 73

\bibitem[{{Spitler} {et~al.}(2014){Spitler}, {Cordes}, {Hessels}, {Lorimer},
  {McLaughlin}, {Chatterjee}, {Crawford}, {Deneva}, {Kaspi}, {Wharton},
  {Allen}, {Bogdanov}, {Brazier}, {Camilo}, {Freire}, {Jenet},
  {Karako-Argaman}, {Knispel}, {Lazarus}, {Lee}, {van Leeuwen}, {Lynch},
  {Ransom}, {Scholz}, {Siemens}, {Stairs}, {Stovall}, {Swiggum},
  {Venkataraman}, {Zhu}, {Aulbert}, \& {Fehrmann}}]{2014ApJ...790..101S}
{Spitler}, L.~G., {et~al.} 2014, \apj, 790, 101

\bibitem[{{Stinebring}(2006)}]{2006ChJAS...6b.204S}
{Stinebring}, D.~R. 2006, Chinese Journal of Astronomy and Astrophysics
  Supplement, 6, 020000

\bibitem[{{Stinebring} {et~al.}(2000){Stinebring}, {Smirnova}, {Hankins},
  {Hovis}, {Kaspi}, {Kempner}, {Myers}, \& {Nice}}]{2000ApJ...539..300S}
{Stinebring}, D.~R., {Smirnova}, T.~V., {Hankins}, T.~H., {Hovis}, J.~S.,
  {Kaspi}, V.~M., {Kempner}, J.~C., {Myers}, E., \& {Nice}, D.~J. 2000, \apj,
  539, 300

\bibitem[{{Thornton} {et~al.}(2013){Thornton}, {Stappers}, {Bailes},
  {Barsdell}, {Bates}, {Bhat}, {Burgay}, {Burke-Spolaor}, {Champion}, {Coster},
  {D'Amico}, {Jameson}, {Johnston}, {Keith}, {Kramer}, {Levin}, {Milia}, {Ng},
  {Possenti}, \& {van Straten}}]{2013Sci...341...53T}
{Thornton}, D., {et~al.} 2013, Science, 341, 53

\end{thebibliography}

\end{document}